\documentclass[iop]{emulateapj}

\usepackage{amsmath}
\usepackage{amssymb}
\usepackage{color}
\usepackage{epic}
\usepackage{epstopdf}
\usepackage{graphicx}
\usepackage{hyperref}
\usepackage{multirow}
\usepackage{natbib}
\usepackage{overpic}

\begin{document}

\title{The Energetics and Lifetimes of Local Radio Active Galactic Nuclei}

\author{Ross J. Turner}
\email{Email: rjturner@utas.edu.au}
\author{Stanislav S. Shabala}
\affil{School of Physical Sciences, Private Bag 37, University of Tasmania, Hobart, TAS, 7001, Australia}

\begin{abstract}

We present a model describing the evolution of Fanaroff-Riley type I and II radio AGN, and the transition between these classes. We quantify galaxy environments using a semi-analytic galaxy formation model, and apply our model to a volume-limited low redshift ($0.03 \leqslant z \leqslant 0.1$) sample of observed AGN to determine the distribution of jet powers and active lifetimes at the present epoch. Radio sources in massive galaxies are found to remain active for longer, spend less time in the quiescent phase, and inject more energy into their hosts than their less massive counterparts. The jet power is independent of the host stellar mass within uncertainties, consistent with maintenance-mode AGN feedback paradigm. The environments of these AGN are in or close to long-term heating-cooling balance. We also examine the properties of high- and low-excitation radio galaxy sub-populations. The HERGs are younger than LERGs by an order of magnitude, whilst their jet powers are greater by a factor of four. The Eddington-scaled accretion rates and jet production efficiencies of these populations are consistent with LERGs being powered by radiatively inefficient advection dominated accretion flows (ADAFs), while HERGs are fed by a radiatively efficient accretion mechanism.

\end{abstract}

\keywords{galaxies: active, jets, evolution---radio continuum: galaxies}

\section{INTRODUCTION}
\label{sec:INTRODUCTION}

The growth of galaxies is closely related to that of the supermassive black holes at their centres \citep{Magorrian+1998, HaringRix+2004, Gultekin+2009}. The activity of this central engine is linked to the evolution of the galaxy through infalling matter accreted onto the black hole powering the jets of the active galactic nuclei (AGN). 
The active nucleus can have a profound impact on its surroundings and the host galaxy evolution, through either radiative or kinetic feedback. The kinetic mode feedback is dominant at the present epoch, whilst radiative feedback is prevalent in the early universe \citep[$z \sim 2$ - $3$;][]{Fabian+2012}. Kinetic feedback has been invoked to explain the lack of star formation in the most massive galaxies \citep{Croton+2006}, and the suppression of cooling flows in the cores of massive clusters \citep{Fabian+2003}. 
The energy input from AGN heats up or expels the surrounding intracluster gas (ICM), decreasing the rate of accretion onto the central black hole until it is shut off completely. Accretion can resume once the gas cools, or is replenished through an interaction or merger with another galaxy \citep{Bahcall+1997}. The increase in accretion rate may eventually lead to the retriggering of the radio jets, completing the AGN feedback cycle. Evidence of the sporadic and intermittent nature of this feedback process is found in observations of rising radio bubbles \citep{Churazov+2001} and sources with multiple radio lobe pairs \citep[e.g.][]{Giovannini+1998, Venturi+2004}. 
Quantifying the typical jet powers of these radio sources and the duration of their active state is crucial to understanding the energetics of AGN feedback.

Radio-loud AGN consist of twin jets emanating from the active nucleus in opposite directions. These jets interact with the surrounding medium, leading to the formation of a diffuse emission region. The radio luminosity of this region, termed a cocoon, arises due to the synchrotron emission of shocked electrons (or positrons) injected from the jet. 
Observationally, these radio sources are separated into two classes based on the distribution of radio surface brightness in their cocoons \citep{FR+1974}. \citeauthor{FR+1974} type II sources have well-defined jet termination shocks located towards the ends of edge-brightened cocoons, whilst type I sources have their regions of highest surface brightness close to the core. 
This morphological dichotomy has been argued to arise due to either the influence of external environmental factors on the jet structure \citep[e.g.][]{Laing+1994,Bicknell+1995, KA+1997}, or by parameters associated with the jet production mechanism \citep[e.g.][]{Meier+2001, Garofalo+2010}. Several radio sources have been observed to exhibit a mixed morphology \citep{KB+2007}, suggesting the FR-I/II dichotomy is at least partly due to the environment. Both FR-I and FR-II radio sources are thought to initially expand supersonically \citep{Marecki+2003}, and it has thus been suggested that FR-II radio sources may evolve into FR-Is when their jets are disrupted \citep{GW+1988, KB+2007}.
There are numerous models in the literature describing the temporal evolution of the powerful FR-II class \citep[e.g.][]{KA+1997, Blundell+1999, KC+2002}, and some models for the FR-I morphological class \citep[e.g.][]{LS+2010}. These models typically make a number of simplifications to remain analytically tractable. External density profiles are usually approximated by a single power law with a constant exponent, inconsistent with X-ray observations of the gas density in clusters \citep[e.g.][]{Vikhlinin+2006}. The separate FR-II and FR-I models also only consider the two limiting cases, where either the ram (for FR-IIs) or external pressure (FR-Is) component of the total pressure dominates.

Properties of AGN central engines can be studied using optical emission lines \citep[see e.g.][]{Hardcastle+2009, Best+2012}. {In this approach, AGN are partitioned into two populations, high- and low-excitation radio galaxies (or HERGs and LERGs), based on either line ratios or line equivalent widths.} These emission line properties are closely correlated with radio source morphological classification, with almost all FR-Is found in LERGs and many, but not all, FR-IIs inhabiting HERGs \citep{Hardcastle+2007, Lin+2010}. 
These two populations are thought to host black holes powered by different accretion flow mechanisms \citep[e.g.][]{Hardcastle+2007, Best+2012}, with LERGs fuelled by an advection dominated accretion flow \citep[ADAF,][]{Narayan+1995} and the radiatively efficient HERGs by a thin \citep{Shakura+1973} or possibly slim \citep{Abramowicz+1988} disk flow. The accretion flow state near a supermassive black hole is largely dependent on its accretion rate, though spin also contributes. Black holes with low Eddington-scaled accretion rates $\dot{m} \ll 0.01$ are thought to be fuelled by ADAFs and those with higher rates of accretion between $\dot{m} = 0.01$ and $0.3$ powered by a thin disk \citep[e.g.][]{Meier+2001, Park+2001}.

In this paper, we develop a hybrid model for both FR-II and FR-I type sources, and the transition between these morphologies (Sections \ref{sec:RADIO SOURCE MODEL} and \ref{sec:RAYLEIGH-TAYLOR MIXING}). We model the evolution of radio sources in gaseous atmospheres derived from a semi-analytic galaxy formation model and apply our model to a local AGN sample to characterise their energetics (Section \ref{sec:ENVIRONMENT AND PARAMETER ESTIMATION}). In Section \ref{sec:RADIO SOURCE MORPHOLOGY}, we examine the relationship between the radio cocoon--ambient gas interaction and radio source morphology. Finally, in Sections \ref{sec:AGN ENERGETICS AND FEEDBACK} and \ref{sec:FUELLING THE MONSTERS} these results are compared to the theoretical expectations for AGN feedback and used to examine the accretion properties of high- and low-excitation radio galaxies.

The $\Lambda \rm CDM$ concordance cosmology with $\Omega_{\rm M} = 0.3$, $\Omega_\Lambda = 0.7$ and $H_0 = 70 \rm\,km \,s^{-1} \,Mpc^{-1}$ is assumed throughout the paper.

\section{RADIO SOURCE MODEL}
\label{sec:RADIO SOURCE MODEL}

In this section, we develop the dynamical model. Our model is largely based on the FR-II model of \citet{KA+1997}, and the pressure-limited expansion model of \citet{LS+2010}. We combine these two limiting cases in a single framework, and include a more complete treatment of the environments into which radio sources expand. The key parameters defined in this section are summarised in Table \ref{tab:modelvars}.

\subsection{Radio Source Dynamics}
\label{sec:Radio Source Dynamics}

\begin{figure*}
\begin{center}
\includegraphics[width=0.48\textwidth]{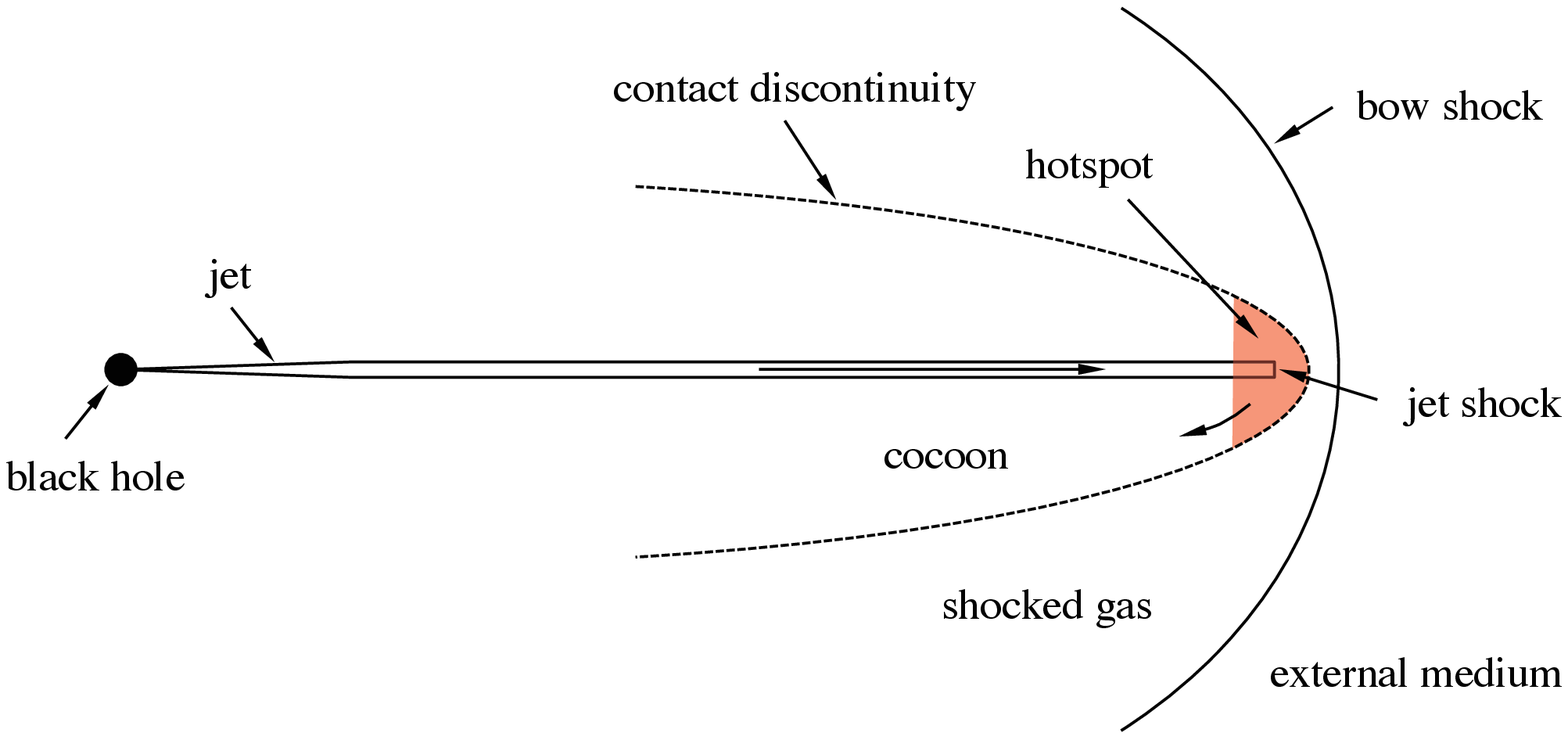}\quad\quad\includegraphics[width=0.48\textwidth]{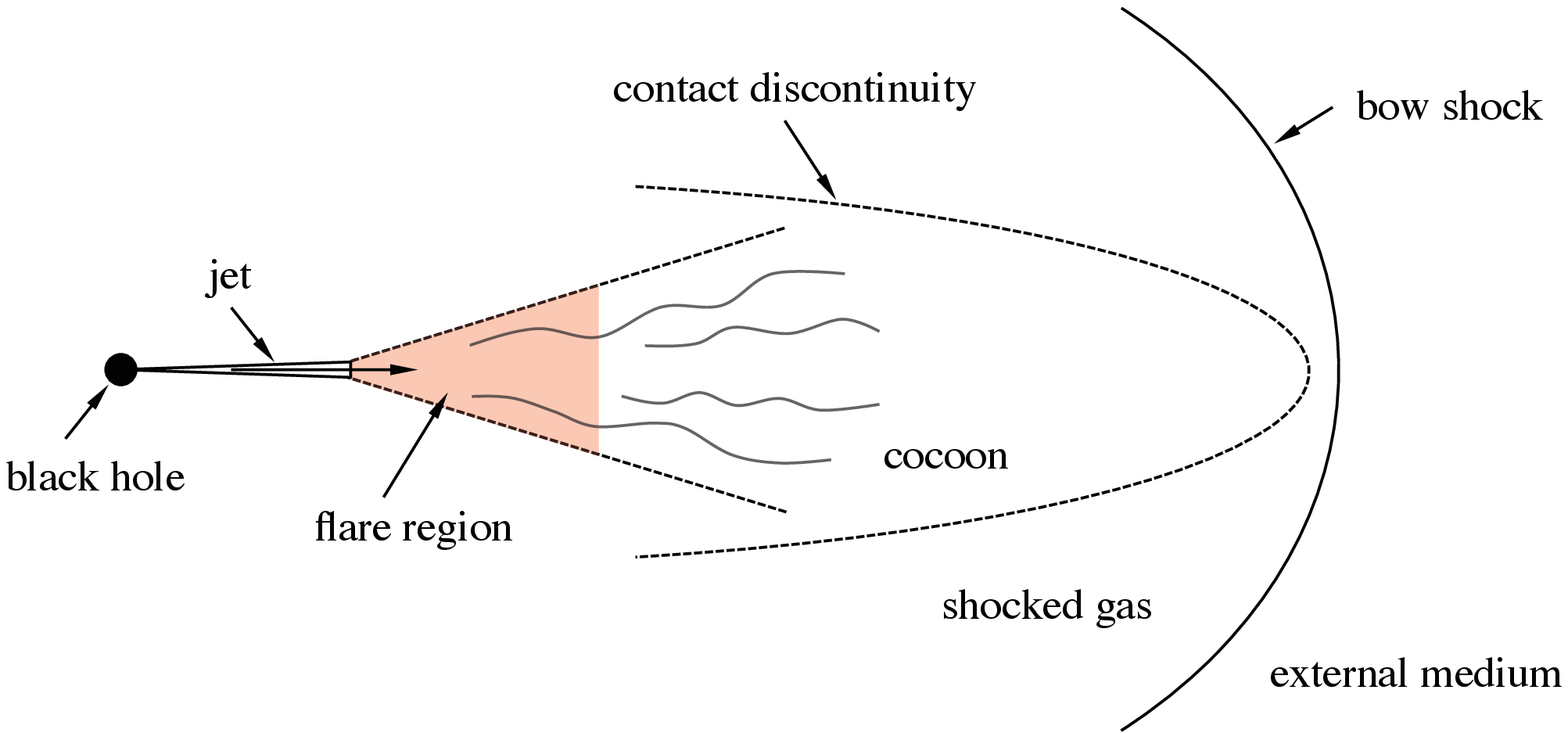} 
\end{center}
\caption{FR-II (left) and FR-I (right) model schematics. Jets and cocoons on only one side of the AGN are shown.}
\label{fig:frSmodel}
\end{figure*}

The FR-II dynamical model comprises a relativistic plasma jet emanating ballistically from the active nucleus with constant opening angle (Figure \ref{fig:frSmodel}a). A reconfinement shock will form further down the jet enabling it to transition from the ballistic regime into one of pressure equilibrium with its cocoon. The radius of the jet will remain constant after this reconfinement shock since the cocoon pressure remains uniform \citep[due to the high sound speed;][]{KA+1997}. The jet terminates at an outwardly moving hotspot in a jet shock with the ram pressure of the jet material distributed over the working surface, generating a bow shock. The pressure exerted on the working surface is balanced by the pressure of the shocked ambient gas which surrounds the hotspot. The hotspot region is overpressured with respect to the rest of the cocoon, and backflow of overpressured plasma inflates the cocoon. The cocoon and bow shock are expected to expand in a self-similar manner \citep{Falle+1991, KA+1997}. The luminosity of the cocoon comes from the radio emission due to synchrotron electrons precessing about the magnetic field lines of the cocoon.

In the FR-I model (Figure \ref{fig:frSmodel}b), a relativistic jet emerges from the central black hole, inflating a cocoon, similar to the FR-II model. Particles will still be accelerated by jet shocks, however, these shocks occur closer to the nucleus for the low-power jets of FR-Is. The region where shocks occur is referred to as the flare region, the FR-I counterpart of the FR-IIs hotspots \citep[e.g.][]{LS+2010}. The diffuse emission region, which extends well beyond the shock, is regarded as the lobe or cocoon in analogy with FR-IIs.

\subsection{Environment assumptions}
\label{sec:Environment assumptions}

We model the expansion of the pressure-inflated radio cocoon in a spherically symmetric external atmosphere with an arbitrary density and temperature profile. The external atmosphere is partitioned into small radial regions in which the external density and temperature are each approximated by a simple power law profile. 
The density profile in each small region is thus approximated by a power law of the form

\begin{equation}
\rho_{\rm x} = \rho_0 \left(\frac{r}{a_0} \right)^{-\beta} = k r^{-\beta} ,
\label{density profile}
\end{equation}

where $r$ is the radial distance from the active nucleus at the centre of the source and the exponent $\beta$ is constant in the region. The density parameter $k$ is defined as $k = \rho_0 {a_0}^\beta$ for some radius $a_0$, with corresponding density $\rho_0$, located in the region approximated by the relevant power law.

The temperature of the external environment in each small region is similarly modelled as

\begin{equation}
\tau_{\rm x} = \left(\frac{\bar{m}}{k_{\rm B}} \right) l r^{-\xi} ,
\label{temp profile}
\end{equation}

where the exponent $\xi$ is constant in the region, $k_{\rm B}$ is the Boltzmann constant, $\bar{m} \sim 0.6 m_{\rm p}$ is the average particle number density for proton mass $m_{\rm p}$, and $l$ is a constant of proportionality termed the temperature parameter\footnote{The $\bar{m}/k_{\rm B}$ constant is introduced in this definition to simplify subsequent equations.}.

\subsection{Transonic dynamical model}
\label{sec:Transonic dynamical model}

Dynamical models of powerful type II radio sources show that the overpressured cocoons expand self-similarly, i.e. the shape of the cocoon remains fixed \citep{Falle+1991, KA+1997}. As the cocoon expansion slows, however, the surface velocity of the inner parts of the cocoon will fall below the local sound speed of the ambient gas. The expansion rate in this subsonic phase \citep[$r \propto t^{1/(3 - \beta)}$;][]{LS+2010} is lower than for supersonic expansion \citep[$r \propto t^{3/(5 - \beta)}$, for realistic density profiles with $\beta < 2$;][]{KA+1997}, and thus the assumption of self-similar growth cannot be extended to this transonic regime. 
We therefore model the cocoon as an ensemble of small volume elements in pressure equilibrium, where the growth of each element is otherwise independent. In the subsonic phase we further consider Rayleigh-Taylor instabilities expanding about the surface of this volume (see Section \ref{sec:RAYLEIGH-TAYLOR MIXING}).

In order to quantify the energetics of the adiabatic expansion of the radio source we need expressions for the volume of the cocoon and its pressure. The evolution of the cocoon on one side of the central active nucleus is modelled explicitly, with the properties of the entire radio source calculated by assuming two identical, anti-parallel cocoons.
Below we derive expressions for the size and luminosity evolution of the radio cocoons.

\subsubsection{Cocoon geometry}

\begin{figure}[b]
\begin{center}
\includegraphics[width=0.42\textwidth]{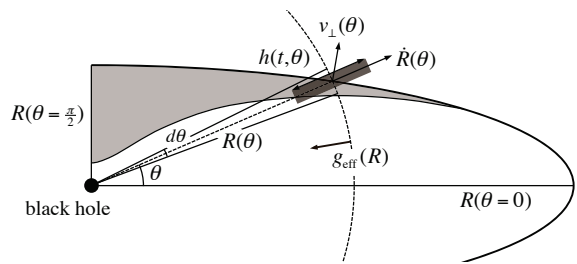} 
\end{center}
\caption{A cross section of the entire cocoon (solid line), with the non-synchrotron emitting, Rayleigh-Taylor mixed region shaded in light grey. Geometric parameters are shown in this schematic for the ensemble element at angle $\theta$. The shaded rectangle of length $h(t,\theta)$ represents the height of the mixing layer. The height of the mixing layer increases towards the cocoon's semi-minor axis where the normal velocity and thus Mach number are lower.}
\label{fig:cocoonmix}
\end{figure}

The radio cocoon of our dynamical model is an ensemble of small angular volume elements in pressure equilibrium. Each element of fixed angular width is assumed to receive a constant fraction of the jet power as the cocoon expands. This assumption yields self-similar expansion when the whole cocoon is in the supersonic phase. The volume of each small angular cocoon element $[\theta - d\theta/2, \theta + d\theta/2)$ is thus related to its radial length $R(\theta)$ by

\begin{equation}
dV(\theta) = \frac{2\pi R^3(\theta)}{3} \sin\theta d\theta ,
\label{delta volume}
\end{equation}

where $\theta$ is the angle between the jet axis and the radial line passing through the volume element (Figure \ref{fig:cocoonmix}). The total volume of the cocoon on one side of the central nucleus is the volume integral of Equation \ref{delta volume} over the domain $\theta \in (0,\tfrac{\pi}{2})$. To provide initial conditions for this model, we assume our cocoon is initially ellipsoidal in shape with a circular cross-section along the jet axis. The ratio of the length of the cocoon on one side of the central active nucleus, $R(\theta = 0)$, to the radius of the cocoon transverse to the jet axis is defined by $A$, the cocoon axis ratio\footnote{Hence for a sphere we have $A = 1$. This axis ratio is related to the aspect ratio $R_{\rm T}$ defined by \citet{KA+1997} as $A = 2R_{\rm T}$.}. 

\subsubsection{Working surface pressure}

The pressure at the working surface of highly supersonic FR-II radio sources is normally calculated using ram pressure arguments or the Rankine-Hugoniot jump conditions for a plane parallel shock \citep[e.g.][]{KA+1997, LS+2010}. For a subsonic FR-I source, pressure due to such expansion is assumed insignificant with the working surface pressure taken to be equal to the external pressure \citep[e.g.][]{LS+2010}. The working surface of the cocoon is defined here as the location where the expanding cocoon and external environment meet.

For a compressible fluid and ignoring external forces, the working surface pressure is related to the external pressure $p_{\rm x}(\theta)$ using the Rankine-Hugoniot jump conditions for a plane parallel shock. The working surface pressure approaches the external pressure as the cocoon expansion speed nears the sound speed, i.e. $M_{\rm b} \rightarrow 1$ where $M_{\rm b}(\theta)$ is the Mach number of the bow shock. In the subsonic regime we do not expect the working surface pressure to fall below the external pressure. The pressure at the working surface should thus equal the external pressure for subsonic expansion; this is the ``pressure-limited expansion'' of \citet{LS+2010}. The general expression for the working surface pressure $p(\theta)$ of each ensemble element in the cocoon is therefore

\begin{equation}
p = 
\begin{cases}
   \frac{2 \Gamma_{\rm x} {M_{\rm b}}^2 - (\Gamma_{\rm x} - 1)}{\Gamma_{\rm x} + 1} p_{\rm x} & \text{for } M_{\rm b} \geqslant 1 \\
   p_{\rm x} & \text{for } M_{\rm b} < 1 ,
  \end{cases}  
\label{hotspot pressure 3}
\end{equation}

where $\Gamma_{\rm x} = \tfrac{5}{3}$ is the adiabatic index of the external medium. This equation can be rewritten in terms of the radial length of the ensemble element $R(\theta)$, and the density and temperature parameters, $k$ and $l$ respectively.
The pressure of the external medium is given by $p_{\rm x} = \rho_{\rm x} \tau_{\rm x} (k_{\rm B}/\bar{m}) = (k l) R^{-(\beta + \xi)}$, where the right hand equality is obtained using Equations \ref{density profile} and \ref{temp profile}. The Mach number of the bow shock is related to the expansion rate of the ensemble element as $M_{\rm b} = v_\perp/c_{\rm x}$, where $v_\perp(\theta) = \zeta \dot{R}/\eta$ is the velocity normal to the surface (see Appendix for derivation and definitions of the dimensionless radius and velocity, $\eta$ and $\zeta$) and $c_{\rm x} (\theta) = \sqrt{\Gamma_{\rm x} k_{\rm B} \tau_{\rm x} (\theta)/\bar{m}}$ is the sound speed. The Mach number can then be written as ${M_{\rm b}} = [R^{\xi} (\zeta \dot{R}/\eta)^2/(\Gamma_{\rm x} l)]^{1/2}$, and hence Equation \ref{hotspot pressure 3} becomes

\begin{equation}
p = 
\begin{cases}
   \frac{2}{\Gamma_{\rm x} + 1} k R^{-\beta} (\zeta \dot{R}/\eta)^2 - \frac{\Gamma_{\rm x} - 1}{\Gamma_{\rm x} + 1} (k l) R^{-(\beta + \xi)} & \text{for } M_{\rm b} \geqslant 1 \\
  (k l) R^{-(\beta + \xi)} & \text{for } M_{\rm b} < 1 .
  \end{cases}
\label{hotspot pressure 4}
\end{equation}

The average pressure inside the cocoon is calculated by weighting the surface pressure of each ensemble element by its volume, since the high sound speed is expected to yield uniform internal pressure.

\begin{table}
\begin{center}
\caption[]{Environment and cocoon geometry parameters.}
\label{tab:modelvars}
\renewcommand{\arraystretch}{1.3}
\setlength{\tabcolsep}{10pt}
\begin{tabular}{cccc}
\hline\hline
\multicolumn{4}{c}{Pressure profile parameters}\\
\hline \vspace{-0.35cm}\\
density parameter&$k$&\multicolumn{2}{c}{\multirow{2}{*}{$\rho_{\rm x} = k r^{-\beta}$}}\\
density exponent&$\beta$&\vspace{0.05cm}\vspace{0.05cm}\\
temperature parameter&$l$&\multicolumn{2}{c}{\multirow{2}{*}{$\tau_{\rm x} = l r^{-\xi}$}}\\
temperature exponent&$\xi$&\vspace{0.035cm}\\
\hline\hline
\multicolumn{4}{c}{Ensemble element parameters}\\
\hline \vspace{-0.35cm}\\
\multicolumn{2}{c}{angle of element from jet axis}&\multicolumn{2}{c}{$\theta$}\\
\multicolumn{2}{c}{fraction of jet power in element}&\multicolumn{2}{c}{$d\lambda$}\\
\multicolumn{2}{c}{volume of ensemble element}&\multicolumn{2}{c}{$dV$}\vspace{0.1cm}\\
\multicolumn{2}{c}{radius of ensemble element}&\multicolumn{2}{c}{$R$}\\
\multicolumn{2}{c}{radial velocity at element surface}&\multicolumn{2}{c}{$v$}
\vspace{0.035cm}\\
\hline\hline
\multicolumn{4}{c}{Dimensionless constants}\\
\hline \vspace{-0.35cm}\\
dimensionless radius&\multicolumn{3}{c}{$\eta = R(\theta)/R(\theta = 0)$}\\
dimensionless velocity&\multicolumn{3}{c}{$\zeta = v_\perp(\theta)/\dot{R}(\theta = 0)$}\vspace{0.035cm}\\
\hline
\end{tabular}
\end{center}
\end{table}

\subsection{Adiabatic expansion}
\label{sec:Adiabatic expansion}

The adiabatic expansion of each angular volume element in the cocoon ensemble is related to the pressure imparted on that element at the surface, $p(\theta)$, its volume $dV(\theta)$ and the jet power $Q$ by

\begin{equation}
\dot{p} \,dV + \Gamma_{\rm c} p \,d\dot{V} = (\Gamma_{\rm c} - 1) Q\,  d\lambda ,
\label{adiabatic expansion}
\end{equation}

where $\Gamma_{\rm c} = \tfrac{5}{3}$ \citep{KDA+1997} is the adiabatic index of the cocoon and the derivatives are with respect to the time $t$. The differential quantity $d\lambda(\theta)$ is the fraction of the jet power injected into the ensemble element $[\theta - d\theta/2, \theta + d\theta/2)$, derived in the Appendix. Note that $Q$ is the kinetic power of one jet, i.e. the total AGN kinetic power is $Q_{\rm tot} = 2Q$.  The source linear size (an observable quantity) is then related to the length of the ensemble element along the jet axis by $D = 2 R(\theta = 0)$. The subsonic expansion phase, however, must be considered separately from the supersonic phase due to the discontinuity in cocoon pressure in Equation \ref{hotspot pressure 4}. We note that the hotspot diameters of typical FR-II sources are approximately 100 times smaller than the linear size of their cocoons \citep{Hardcastle+1998}. The energy stored in these high pressure hotspots can therefore be neglected from Equation \ref{adiabatic expansion} since it is small compared to that stored in the rest of the cocoon.

\subsubsection{Supersonic expansion}
\label{sec:Supersonic expansion}

We first consider the supersonic phase expansion of ensemble elements in the cocoon. Substituting the relevant expressions for the pressure and volume of each element into Equation \ref{adiabatic expansion} yields a second order differential equation, which can not be solved analytically in general. We therefore adopt a numerical scheme using a fourth order Runge-Kutta method by rewriting the differential equation as a system of two first order ODEs. Thus for each ensemble element in the cocoon we must solve the following system of equations:

\begin{equation}
\begin{split}
\dot{R} &= v \\
\dot{v} &= \frac{3 (\Gamma_{\rm x} + 1)(\Gamma_{\rm c} - 1) Q R^{\beta - 3} d\lambda}{8 \pi v (\zeta/\eta)^2 k \sin\theta d\theta} + \frac{(\beta - 3\Gamma_{\rm c}) v^2}{2 R} \\
&\quad\quad + \frac{(\Gamma_{\rm x} - 1) (3 \Gamma_{\rm c} - \beta - \xi) l}{4 R^{\xi + 1} (\zeta/\eta)^2} ,
\end{split}
\label{supersonic system}
\end{equation}

where $v(\theta)$ is the radial velocity at which the cocoon expands along each angle $\theta$. The first of these equations thus gives the velocity of the expanding cocoon and the second its acceleration.

\subsubsection{Subsonic expansion}
\label{sec:Subsonic expansion}

We now examine the expansion of ensemble elements in the subsonic phase. 
Substituting the expressions for the subsonic pressure (Equation \ref{hotspot pressure 4}) and volume (Equation \ref{delta volume}) into Equation \ref{adiabatic expansion} yields a first order differential equation, which can be solved analytically. The solution to this differential equation is of the form $R \propto t^{1/(3 - \beta - \xi)}$. However, this expression can only be used to model the evolution of a radio source that is subsonic for its entire lifetime. In particular, the ensemble element radius predicted by the subsonic solution at the sonic boundary differs from that of the supersonic phase (Equation \ref{supersonic system}), because the subsonic solution ignores the previous (supersonic) expansion history of the radio cocoon. The expansion in the subsonic phase must therefore be solved numerically, using the supersonic phase solution as an initial condition. We thus rewrite our analytic solution as a system of differential equations,

\begin{equation}
\begin{split}
\dot{R} &= v \\
\dot{v} &= \frac{(\beta + \xi - 2) v^2}{R} ,
\end{split}
\label{subsonic system}
\end{equation}

where again the first equation gives the velocity and the second the acceleration for each ensemble element.

\subsubsection{Conditions of continuity}
\label{sec:Conditions of continuity}

The differential equations derived in the previous sections are only applicable in an environment with constant density and temperature exponents (i.e. constant $\beta$ and $\xi$). 
For general environments, in practice we approximate the density and temperature profiles as a series of power law segments (typically 100 segments per source), and ensure the continuity of ensemble element sizes and pressures across the region boundaries.
These requirements are met by using the radius $R$ and velocity $\dot{R}$ in each region as initial conditions for the next one.

\begin{figure}
\begin{center}
\includegraphics[width=0.45\textwidth]{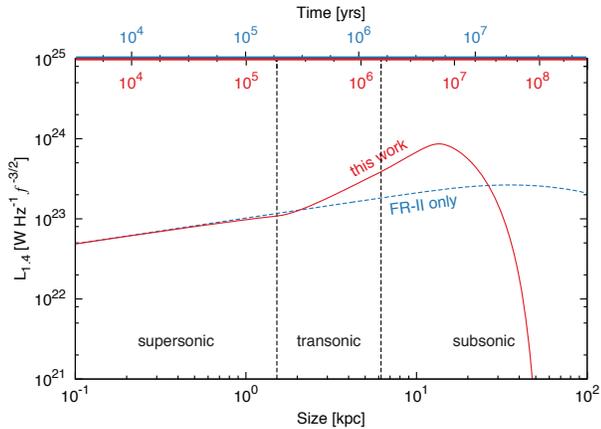}
\end{center}
\caption{{Luminosity--size profile (or $L\,$--$\,D$ track; red curve) for a $10^{11} \rm\, M_\odot$ stellar mass AGN host galaxy with a jet power of $Q_{\rm tot} = 10^{35.5} {\rm\, W}$ in a typical environment. The transonic phase occurs when some parts of the cocoon are expanding subsonically, while others are still supersonic. The fall in luminosity beyond $10 \rm\, Myr$ is due to the Rayleigh-Taylor mixing of the cocoon with its environment. The dotted blue line shows the evolution predicted by our model using the supersonic solution throughout the source lifetime. Note the different time axes for the two models {(lower axis for our model, and upper axis for the FR-II only model)}. These models yield very different sizes and luminosities at ages exceeding $1 \rm\, Myr$.}}
\label{fig:pdtrack}
\end{figure}

\subsection{Radio source luminosity}
\label{sec:Radio source luminosity}

In order to observationally constrain the properties of radio AGN, we need to model the radio luminosity due to the synchrotron emission from the relativistic electrons in AGN cocoons. Synchrotron-emitting electrons will lose energy through adiabatic, synchrotron and inverse Compton losses. We follow the formalism of \citet{KDA+1997}, modifying their Equation 16 to include our multi-power law pressure profile. The total cocoon luminosity (i.e. both lobes) at frequency $\nu$ is then given by

\begin{equation}
\begin{split}
L_\nu = \int_0^t \frac{4 \sigma_{\rm T} c q_{\rm B_p}}{3 \nu (q_{\rm B_p} + 1)} (Q f^{-3/2}) n_0 \gamma {\gamma_{\rm i}}^{2 - s} \quad\quad\quad\quad \\
A^{2(1 - \Gamma_{\rm c})/\Gamma_{\rm c}} 
\left(\frac{V_{\rm e}(t)}{V_{\rm e}(t_{\rm i})} \right)^{-(\Gamma_{\rm B} + 1/3)} dt_{\rm i} ,
\end{split}
\label{power}
\end{equation}

where $\sigma_{\rm T}$ is the Thompson cross-section, $c$ is the speed of light, $\Gamma_{\rm B} = 5/3$ is the adiabatic index of the magnetic field, {$s = 2.14$ is the electron energy injection index} and $q_{\rm Bp} = (s + 1)/4$ \citep{KDA+1997}. Here, $V_{\rm e}$ is the volume of an electron ensemble where $\gamma$ is the Lorentz factor of an electron emitting radiation predominantly at frequency $\nu$, whilst $\gamma_{\rm i}$ are the Lorentz factors at the times $t_{\rm i}$ when the electrons are initially injected into the radio cocoon. This Lorentz factor $\gamma_{\rm i}$ is calculated iteratively (over each power law) using Equation 10 of \citet{KDA+1997}, and finally used to determine the number density coefficient $n_0$ defined in their Equation 8. 

{We note that there are significant uncertainties in the estimation of the radio source luminosity at a given jet power since the physics of these objects is not fully understood. \citet{Willott+1999} described these uncertainties using a single factor $f$, which is observationally constrained to lie between $\sim 1$ and $20$. In Section \ref{sec:RADIO SOURCE MORPHOLOGY}, we show that our model is consistent with jet powers estimated from X-ray cavity measurements \citep{Heckman+2014} for $f \sim 5$. This factor flows through directly from our jet powers into our measures of the AGN energetics, but also subtly affects our source age estimates. 
For an uncertainty factor of $f = 5$, the estimated source age at a given size and luminosity is approximately $0.3 \rm\, dex$ lower than that calculated using $f = 1$. That is, the source age decreases as roughly $f^{-1/2}$ in agreement with Equations 4 and 5 of \citet{KA+1997} for supersonic expansion. {In this work, we calculate our results using $f = 1$ but retain this uncertainty factor in equations so these results can be readily scaled for any realistic choice of this variable, e.g. $f \lesssim 5$.}}


{A typical luminosity--size track produced by our model is shown in Figure \ref{fig:pdtrack}. Our new model predicts a very different temporal evolution of AGN size and luminosity compared to the standard FR-II models. This has a large effect on the derived jet powers and ages of observed AGN. 
The luminosity increases initially in the core regions where the density of the host environment is roughly constant (with radius) before decreasing as the density profile steepens towards $\beta = 2$. This behaviour is similar to that found in 3D MHD simulations of radio jets in a range of environments \citep{Hardcastle+2013}. By contrast, dynamical models which assume a single $\beta \sim 2$ power law profile \citep[e.g.][not shown in figure]{KDA+1997, Blundell+1999} predict the luminosity decreases monotonically with increasing size. Our model also predicts the source luminosity will rise suddenly upon entering the transonic expansion phase. This increase occurs since the cocoon pressure equilibrates with that of the host environment rather than continuing to drop off at the rate predicted by the FR-II model. \citet{LS+2010} find a similar increase in luminosity using their pressure-limited expansion model for FR-Is. The fall in luminosity beyond $10 \rm\, Myr$ is due to the Rayleigh-Taylor mixing of the cocoon with the host environment as detailed in Section \ref{sec:RAYLEIGH-TAYLOR MIXING}.}

\section{RAYLEIGH-TAYLOR MIXING}
\label{sec:RAYLEIGH-TAYLOR MIXING}

As the radio source expansion slows down, the cocoon becomes susceptible to the Rayleigh-Taylor instability. Different parts of the cocoon become subsonic with respect to the ambient medium at different times. 
As the cocoon plasma becomes entrained by this surrounding medium, the high energy synchrotron electrons in the cocoon will be collisionally reduced to the ambient energy level of the much denser surrounding atmosphere. The radio emission from these mixing regions is thus significantly reduced; here we assume it to be zero. The fluid mechanics of this situation are considered here in order to estimate the volume of the cocoon that remains unmixed, and thus visible, as the cocoon ages. Radio luminosities estimated using the equations of the previous sections will be modified by this visible fraction to account for the Rayleigh-Taylor mixing. 

The expanding surface of the underdense cocoon will become Rayleigh-Taylor unstable \citep{Rayleigh+1883, Taylor+1950} as it enters the subsonic phase. The thickness of the Rayleigh-Taylor mixing layer can be calculated based on the late time self-similar growth phase \citep{Fermi+1953}\footnote{The initial exponential growth phase is insignificant in comparison to the self-similar growth. Using linear stability theory of \citet{Rayleigh+1883} we found that a delay of 10 $e$-folding times yielded a less than $10\%$ variation in the unmixed volume and hence luminosity.}. The rate of expansion of this layer can be shown \citep{Cook+2004, Ristorcelli+2004} to be described by

\begin{equation}
\frac{dh}{dt} = 2 \left[\frac{\kappa (\rho_{\rm x} - \rho_{\rm coc}) g_{\rm eff} h}{(\rho_{\rm x} + \rho_{\rm coc})} \right]^{1/2} .
\label{mixing ODE}
\end{equation}

Here $h$ is the thickness of the mixing layer (see Figure \ref{fig:cocoonmix}), $\rho_{\rm x}$ and $\rho_{\rm coc}$ are the densities of the external medium and cocoon respectively, and $\kappa$ is a dimensionless growth parameter. Most lab-based experiments have returned values for $\kappa$ in the range $0.03 < \kappa < 0.07$ \citep[e.g.][]{Dimonte+2004}. The effective gravitational acceleration, $g_{\rm eff}$, is the sum of the gravitational potential of the galaxy, $g_{\rm grav}$, and the acceleration of the cocoon, $g_{\rm coc} = - \dot{v}$ (with $\dot{v}$ as defined in Equation \ref{subsonic system} for the subsonic phase). The gravitational acceleration on the cocoon is given by $g_{\rm grav} = {c_{\rm x}}^2 (\beta/\Gamma_{\rm x} R)$, where $c_{\rm x}$ is the sound speed of the ambient medium with adiabatic index $\Gamma_{\rm x}$, and $\beta$ is the exponent of the power law density profile just outside the cocoon \citep[e.g.][]{Shabala+2009a}. 

A general expression for the thickness of the mixing layer, for constant $\kappa$ but with a time-dependent $g_{\rm eff}$, is then given by 

\begin{equation}
h(t) = \left(\kappa^{1/2} \int_{t_0}^t {g_{\rm eff}}^{1/2}(t) dt + {h_0}^{1/2} \right)^2 .
\label{mixing thickness 2}
\end{equation}

The initial conditions $t_0$ and $h_0 \sim 0$ in this equation are the time at which the flow first enters the self-similar growth phase and the corresponding thickness of the mixing layer. The integral in this equation can be solved numerically by considering small regions over which $g_{\rm eff}(t)$ is approximately constant. For large mixing times it can be assumed that $h_0 = 0$.

The cocoon surface can only become Rayleigh-Taylor unstable some time after the corresponding ensemble element has entered the subsonic expansion phase.
This ensemble element will thereafter entrain material from the external medium in the mixing region. The thickness of the mixing region will grow along the radial axis\footnote{{The same result is obtained if we instead assume that the mixing layer grows normal to the cocoon surface with the effective gravity reduced by the angle $\theta$ between the radial axis and the normal (i.e. $g_{\rm eff \perp} = g_{\rm eff} \cos \theta$). For large times $t \gg t_0$, the mixing layer thickness is proportional to $g_{\rm eff}$, and thus the thickness along the radial axis is $h_\perp(t)/\cos \theta = \kappa g_{\rm eff \perp}/\cos \theta = \kappa g_{\rm eff} = h(t)$.}}, as described by Equation \ref{mixing thickness 2}. The mixing layer is assumed to move with the working surface, expanding equally inward and outward from this equilibrium radius. The visible radius of the cocoon $R_{\rm v}$ at a given angle $\theta$ at each time step is thus $R_{\rm v}(\theta) = R(\theta) - \tfrac{1}{2} h(t, \theta)$. The visible, unmixed volume of the cocoon element is thus $dV_{\rm v}(\theta) = \tfrac{2}{3}\pi {R_{\rm v}}^3(\theta) \sin\theta d\theta$ (c.f. Equation \ref{delta volume}). The fraction of this total cocoon volume that is unmixed and thus visible through radio emission is

\begin{equation}
\mathcal{F} = \frac{\int dV_{\rm v}(\theta)}{\int dV(\theta)} = \frac{\int {R_{\rm v}}^3(\theta) \sin\theta d\theta}{\int R^3(\theta) \sin\theta d\theta} .
\end{equation}

The radio source luminosity calculated in Section \ref{sec:RADIO SOURCE MODEL} is reduced by this factor to yield the luminosity from the non-mixed, synchrotron emitting part of the cocoon, $L_{\nu \rm v} = \mathcal{F} L_\nu$. {We note that this equation assumes uniform luminosity through the unmixed regions of the cocoon. tThis assumption may not strictly be valid in weak FR-I type sources where the luminosity is often dominated by the core.}

\begin{figure}
\begin{center}
\includegraphics[width=0.445\textwidth]{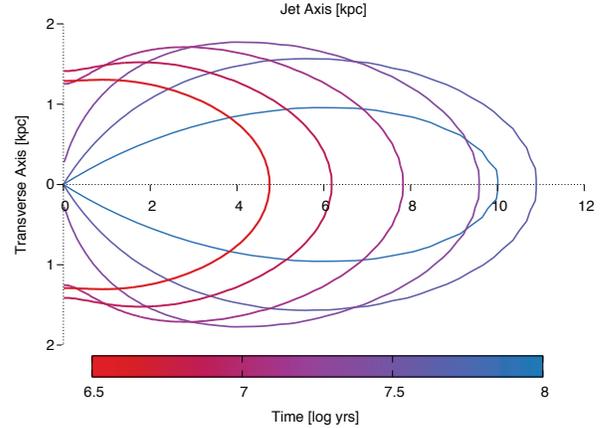} 
\end{center}
\caption{Visible cross-section for a radio source with $Q_{\rm tot} = 10^{35.5} {\rm\, W}$ total jet power and initial axis ratio of $A = 2$, expanding into a host galaxy with typical external pressure profiles described in Section \ref{sec:External pressure profile}.
{The contours on this figure correspond to source ages of $3$, $6$, $12$, $25$, $50$ and $100 \rm\, Myrs$}. The cocoon becomes subsonic and begins to entrain thermal material due to the Rayleigh-Taylor instability, first along the transverse axis (at $\sim 0.2 \rm\, Myrs$), and later along the jet axis (at $\sim 15 \rm\, Myrs$). This Rayleigh-Taylor mixing changes the shape of the visible cocoon, eventually causing its length to \textit{decrease} with time, after $\sim 40 \rm\, Myrs$.}
\label{fig:recession}
\end{figure}

Simulated cross-sections of the visible, unmixed cocoon of a radio source are shown in Figure \ref{fig:recession}. Rayleigh-Taylor instabilities affect the slow-expanding narrow part of the radio cocoon first, giving the cocoon a characteristic hourglass shape observed in FR-II sources such as Cygnus A. The Rayleigh-Taylor mixing may also explain the observed pinching and eventual separation of radio bubbles from their active nucleus \citep[e.g. M87;][]{Forman+2007}. In Section \ref{sec:RADIO SOURCE MORPHOLOGY}, we explore further the relationship between cocoon entrainment and observed morphology.

\section{ENVIRONMENT AND PARAMETER ESTIMATION}
\label{sec:ENVIRONMENT AND PARAMETER ESTIMATION}

In this section, we describe the methodology by which we use our transonic dynamical model to constrain the physical properties of observed AGN. The intrinsic AGN parameters, in particular their jet powers and source ages, are determined for each radio source based on their observed size and luminosity. We quantify radio source environments using semi-analytic galaxy formation and evolution models.

\subsection{Local radio-loud AGN sample}
\label{sec:Local radio-loud AGN sample}

\citet{Shabala+2008} constructed a complete, volume-limited sample of radio AGN at $0.03 \leqslant z \leqslant 0.1$ by cross-matching two VLA radio surveys at 1.4 GHz, the \emph{NRAO VLA Sky Survey} \citep[NVSS;][]{Condon+1998} and \emph{Faint Images of the Radio Sky at Twenty-Centimeters} \citep[FIRST;][]{Becker+1995}, and the optical SDSS Data Release 2. We further pair this sample with the \citet{Yang+2007} galaxy clustering catalogue to obtain an estimate of the galaxy stellar mass and the mass of the halo associated with the cluster. The masses of the central black holes are estimated using the \citet{Gultekin+2009} determination of the $M_{\rm BH\,}$-- $\sigma_\star$ relation for elliptical galaxies, where $\sigma_\star$ is the stellar velocity dispersion. This yields a sample of 615 radio-loud AGN with stellar and halo masses. 

\citet{Best+2012} analysed a similar (but larger) sample of AGN, and separated these into high- and low-excitation radio AGN. Cross-matching our catalogue with theirs, we obtain nine HERG and 336 LERG classifications, with the remaining 270 galaxies unclassified. Both populations are mostly of elliptical galaxy morphology (HERGs $88\%$ and LERGs $93\%$), and we hence adopt environments of simulated elliptical galaxies here (see Section \ref{sec:Hot gas density}).
The emission line fluxes of the radio source host galaxies are taken from the value-added MPA-JHU catalogues \citep{Tremonti+2004}. In particular, we use the [O\,III] 5007 line to calculate the accretion rate of our AGN. 

\subsection{Radio AGN environments}
\label{sec:Radio AGN environments}

\subsubsection{Hot gas density}
\label{sec:Hot gas density}

Environments into which radio sources expand play an important role in determining the observed AGN properties. X-ray observations are not available for the majority of our sample, and in any case would be biased to rich environments. We therefore turn to the semi-analytic galaxy evolution (SAGE) model of \citet{Croton+2006} to estimate realistic values for the total hot gas mass based on the mass of the cluster halo or galaxy sub-halo.
The semi-analytic prescription for the growth of galaxies and their central supermassive black holes is implemented on top of the output of the Millennium N-body simulation \citep{Springel+2005}, and is available from the \emph{Theoretical Astronomical Observatory}\footnote{\url{https://tao.asvo.org.au/tao/}}. 
In this work we use a subset of the mock galaxy catalogue spanning a simulation volume $30h^{-1} \rm\, Mpc$ on the side. The mock galaxy catalogue consists of both elliptical and spiral galaxy populations. Radio-loud AGN in our sample are preferentially hosted by massive elliptical galaxies \citep[Section \ref{sec:Local radio-loud AGN sample}, see also][]{Best+2005b}, and so we only consider these galaxies in quantifying radio source environments.

We compare the hot gas masses obtained from SAGE to observations of the gas mass in clusters \citep{McGaugh+2010, Gonzalez+2013}. The observed gas masses in these $>10^{13.5} \rm\, M_\odot$ halo mass clusters are typically a factor of two lower than predicted by SAGE. We therefore lower the SAGE gas masses by this factor when using our model to estimate the physical parameters of AGN. In Section \ref{sec:Parameter estimation}, we investigate the sensitivity of our results to this choice of hot gas masses.

\begin{figure}
\begin{center}
\includegraphics[width=0.43\textwidth]{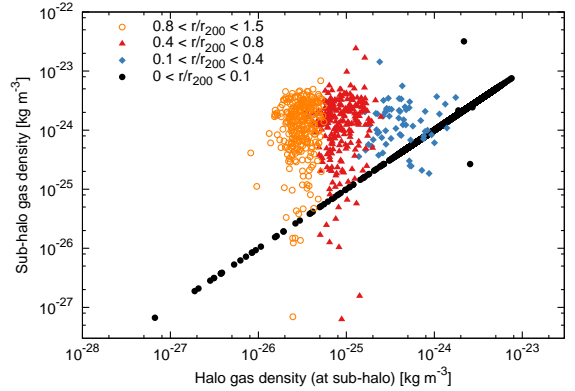} 
\end{center}
\caption{A comparison of main and sub-halo hot gas densities for the galaxies in our sample. The distance of these galaxies from the centre of their group or cluster (scaled by the virial radius) is indicated by the point colour. The straight line of points corresponds to brightest cluster galaxies whose sub-haloes are associated with the majority of the gas in their cluster.}
\label{fig:halodensity}
\end{figure}

An interesting question is whether for AGN that are not hosted by brightest cluster galaxies (BCGs), the relevant hot gas mass is that of the host galaxy sub-halo, or of the whole cluster.
We fit a typical density profile (see Section \ref{sec:External pressure profile}) to the simulated gas mass and virial radius of such haloes. The gas density contribution at a typical radio source size of $10 \rm\, kpc$ from the active nucleus is modelled for both the cluster halo and sub-halo using simulated galaxies in SAGE with comparable stellar masses to the bulk of our sample (i.e. $>10^{10.5} \rm\,M_\odot$). These two contributions to the gas density are plotted in Figure \ref{fig:halodensity}, and grouped by the radial location of the host galaxy within the cluster halo, $r/r_{200}$. Approximately $80\%$ of these are BCGs (black points in the figure), but of the remainder, less than $2\%$ have environments at $10 \rm\, kpc$ with a greater gas density contribution from the cluster halo profile than the host sub-halo. Subsequently we therefore model the external environment using the simulated hot gas density of the host galaxy sub-halo.

\subsubsection{Gas temperature}
\label{sec:Gas temperature}

The temperature of the hot gas in clusters is known to scale with the halo mass as $T \propto {M_{200}}^{0.67 \pm 0.06}$ \citep[e.g.][]{Vikhlinin+2006}, in line with the virial temperature of the halo. However, the gas temperatures in this relationship exclude the central cooling cores and do not include any recent heating from AGN. \citet{OSullivan+2001} compiled a catalogue of 401 early-type galaxies, including 13 AGN hosts, with X-ray luminosity observations as a function of the $B$-band optical luminosity. 
We convert these variables to temperature and halo mass \citep{Bell+2003} to obtain the dependence $T \propto {M_{200}}^{0.27\pm0.24}$. If we include only the ten AGN with haloes in the mass range of our sample we find no significant relationship ($T \propto {M_{200}}^{0.15\pm0.38}$). The temperature of the AGN environment may therefore have only a weak dependence on the mass of its cluster halo. At the median halo mass of our sample, the best fit temperature based on the X-ray luminosity of the AGN is $10^{\,7.0\pm0.3} \rm\, K$.
In this paper we present results that assume this constant mean gas temperature. We note that our findings are relatively insensitive to the choice of temperature profile, since very few of our AGN hosted by low-mass galaxies are expected to be in the subsonic expansion phase where the effects of ambient pressure may be important.

\subsubsection{External pressure profile}
\label{sec:External pressure profile}

The radial dependence of the host gas density and temperature, which together form the overall pressure profile, is modelled using the cluster observations of \citet{Vikhlinin+2006}. Their sample comprises low redshift ($z < 0.3$) clusters with halo masses around $10^{14} \rm\, M_\odot$. The density profiles of these clusters are observed to have very similar shapes, when scaled to the core density and virial radius of the cluster. The temperature of the clusters is similarly constrained, but show far less variability over the radius of the cluster than the density (i.e. $|\xi| \ll |\beta|$, using parameters in Table \ref{tab:modelvars}). The mean gas temperature (Section \ref{sec:Gas temperature}) is therefore adopted for all radii, with $\xi$ taken as zero. 

The gas density profile assumed in this work is a simplified form of the double-$\beta$ profile derived by \citet[][their Equation 3]{Vikhlinin+2006} for a sample of 13 local clusters. Only six of their thirteen clusters need a second, core region $\beta$ profile in their fit, and as such we have chosen to neglect this term here to reduce the degrees of freedom. Our simplified hot gas density profile is thus

\begin{equation}
\rho(r) = {\rho_0} \left(\frac{(r/a_0)^{-\alpha}}{(1+r^2/{a_0}^2)^{3\beta'-\alpha/2}}\frac{1}{(1+r^{\gamma'}/{r_{\rm s}}^{\gamma'})^{\varepsilon/\gamma'}} \right)^{1/2} ,
\label{dense}
\end{equation}

where $\rho_0$ is the mass density in the core, $a_0$ is the core radius and $r_{\rm s} \gtrsim 0.3r_{200}$ is the radius at which the profile steepens. The exponents $\alpha$, $\beta'$, $\gamma'$ and $\varepsilon$ set the slope of the profile, whilst $a_0$ and $r_{\rm s}$ are the core radius and the radius at which the double-$\beta$ profile steepens. The means and standard deviations we assume for these variables in this work are summarised in Table \ref{tab:densevars}.
In our radio source modelling, this continuous density profile is approximated by a series of $\sim 100$ power laws (see Equation \ref{density profile}), conserving the total gas mass $M_{\rm gas}$ of the sub-halo within $r_{200}$ as simulated using SAGE for a given host stellar mass. 
In Section \ref{sec:Robustness to environment}, we show that the derived physical parameters for observed radio sources are relatively insensitive to realistic changes in these external pressure profiles.

\begin{table}
\begin{center}
\caption[]{Ambient pressure profile variables.}
\label{tab:densevars}
\renewcommand{\arraystretch}{1.3}
\setlength{\tabcolsep}{10pt}
\begin{tabular}{cccc}
\hline\hline
\multicolumn{2}{c}{Fixed parameter}&Value&\\
\hline \vspace{-0.35cm}\\
initial axis ratio&$A$&$2$&\\
\multirow{2}{*}{slope (steepening)}&$\varepsilon$&$3$&\\
&$\gamma'$&$3$&\\
steepening radius&$r_{\rm s}/r_{200}$&$0.7$&\vspace{0.035cm}\\
\hline\hline
\multicolumn{2}{c}{Sampled parameter}&Mean&$\sigma$\\
\hline \vspace{-0.35cm}\\
slope (core)&$\alpha$&$1.7$&$0.3$\\
core radius&$a_0/r_{200}$&$0.08$&$0.03$\\
slope&$\beta'$&$0.6$&$0.1$\\
temperature&$\log_{10}\tau_{\rm x}$&7.0&0.3\vspace{0.1cm}\\
gas mass&$M_{\rm gas}$&\multicolumn{2}{c}{\multirow{2}{*}{mock galaxy data}}\\
virial radius&$r_{200}$&\multicolumn{2}{c}{}\vspace{0.035cm}\\
\hline
\end{tabular}
\end{center}
\end{table}

\subsection{Jet powers and ages}
\label{sec:Parameter estimation}

The external pressure profile $p_{\rm x} (r)$ of each observed galaxy is modelled by a set of 128 profiles based on the \citet{Vikhlinin+2006} density and temperature profiles with parameters given in Table \ref{tab:densevars}. For each profile, we sample the gas mass and virial radius distributions from the simulated SAGE galaxies, at the stellar mass of interest, and normal distributions (shown in Table \ref{tab:densevars}) on parameters describing the shape of the density and temperature profiles. We simulate radio source evolution for a range of jet powers $Q_{\rm tot}$ and ages $t_{\rm age}$ in each pressure profile. The goodness of fit of the simulated luminosity--size profile (or $L\,$--$\,D$ track) for each such environment to the observed galaxy size $D_{\rm obs}$ and luminosity $L_{\rm obs}$ (in log-space) is assessed using the $\chi^2$ statistic,

\begin{equation}
\chi^2 = \left(\frac{D_{\rm pred} - D_{\rm obs}}{\sigma_{\rm D}} \right)^2 + \left(\frac{L_{\rm pred} - L_{\rm obs}}{\sigma_{\rm L}} \right)^2 ,
\end{equation}

where $D_{\rm pred}$ and $L_{\rm pred}$ are the corresponding parameters predicted by our model. Here the uncertainty due to the model dominates and is taken as $\sigma_{\rm D} = \sigma_{\rm L} = 0.3 \rm\, dex$. The likelihood of each simulation matching the observations is calculated as $P_{\rm pred} = e^{-\chi^2/2}$, yielding a joint probability distribution $P(Q_{\rm tot}, t_{\rm age}, p_{\rm x})$. The marginalised probability for the $n$-th value of the simulated jet power $Q_{\rm tot\, n}$ matching the observations is then

\begin{figure}
\begin{center}
\includegraphics[width=0.465\textwidth]{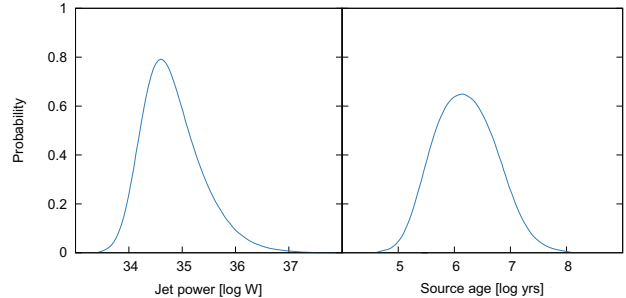} 
\end{center}
\caption{Marginalised, normalised one-dimensional probability distributions for the jet power (left) and source age (right) for typical AGN in our sample.}
\label{fig:marginalised}
\end{figure}

\begin{equation}
P(Q_{\rm tot\, n}) = \sum_{i=1} \sum_{j=1} P(Q_{\rm tot\, n}, t_{\rm age\, i}, p_{\rm x\, j}) ,
\label{marginal}
\end{equation}

with an equivalent expression for the source age,

\begin{equation}
P(t_{\rm age\, n}) = \sum_{i=1} \sum_{j=1} P(Q_{\rm tot\, i}, t_{\rm age\, n}, p_{\rm x\, j}) .
\label{marginal2}
\end{equation}

These probability distributions are approximately log-normal (Figure \ref{fig:marginalised}). The best estimates for the jet power and source age are thus taken as the peaks of the relevant marginal distributions, with the uncertainties taken as the FWHM. We repeat this procedure for each of the 615 AGN in our sample.

\subsubsection{Robustness to jet power and age priors}

The joint probability distribution is initially calculated assuming a flat prior (in log-space) for the jet power and source age, and a log-normal distribution for the host gas density. The estimated distributions of these parameters are coupled (e.g. both old, high jet power, and young, low jet power sources will appear faint and are thus more likely to be undetectable in our radio surveys). We therefore relaxed the assumption of log-uniform priors, instead adopting the estimated distributions (from the first pass) of jet powers and ages in the marginalisation process. The results are robust to these changes in the priors on jet power and source age.

\subsubsection{Robustness to environment}
\label{sec:Robustness to environment}

We further examine the reliability of our jet power and source age estimates by assessing their sensitivity to the host gas density prior. The jet powers and source ages are recalculated for gas density priors with a factor of two shift (i.e. $0.3 \rm\, dex$) in the mean density of the distribution at each stellar mass.
The shifts in the estimated jet power and source age distributions are shown in Figures \ref{fig:QjetDistMass} through \ref{fig:ESDistMass}. The higher pressure prior yields lower jet powers, in particular for our lower mass hosts, and older sources in all galaxies. This is expected since higher external pressures yield higher luminosities \citep[][Equation 16]{KDA+1997} requiring a lower jet power to match observations. The source age is necessarily greater due to the increased resistance to expansion. Conversely, the lower pressure prior corresponds to higher jet powers and younger sources. These factor of two shifts in the mean of the external pressure distribution result in a less than $0.15 \rm\, dex$ variation in the median jet power and source age. These variations are both much less than the intrinsic width in the derived jet power and source age distributions ($0.6$ and $0.9 \rm\, dex$ respectively). We conclude that the jet power and source age parameter estimates are therefore robust to the moderate uncertainties in the distribution of the pressure profiles. We estimate these parameters are accurate to better than a factor of 1.5 (in a statistical sense). We note that our assumption of a cluster-like pressure profile may not be applicable for low-mass AGN hosts. However, the assumed gas masses (from SAGE) are consistent with observed galaxy properties in a statistical sense, validating our approach.

\subsection{Axis ratio}
\label{sec:Axis ratio}

The initial supersonic phase value of the axis ratio, $A$ (this is the ratio of lobe length and radius), is a free parameter in our model. We adopt a value of $A = 2$ here for all sources, noting that $A$ increases once the cocoon enters the transonic phase. Our model predicts axis ratios in the range 2 - 8, broadly consistent with the observed radio AGN population \citep{Mullin+2008}. We defer a detailed investigation of cocoon shape to a future paper.

\section{RADIO SOURCE MORPHOLOGY}
\label{sec:RADIO SOURCE MORPHOLOGY}

The \citeauthor{FR+1974} morphological dichotomy can be at least partly explained by the influence of external environmental factors on the jet structure \citep[e.g.][]{Laing+1994,Bicknell+1995, KA+1997}.
In particular, shredding of the subsonic cocoon by Rayleigh-Taylor mixing of the external medium makes the jet more liable to Kelvin-Helmholtz instabilities, leading to an FR-I like morphology. 
Here, we test whether the cocoon expansion speed predicted by our dynamical model is related to radio source morphology. In particular, we hypothesise that sources expanding supersonically will appear as FR-IIs, and those expanding subsonically as FR-Is.

In our model, morphology is determined by calculating the fraction of their lives that radio sources of a given jet power and stellar mass spend in the subsonic and supersonic expansion phases. The source active lifetimes are constrained by fitting a multivariate regression to the jet power and stellar mass estimates we obtained for our sample. Because the cocoon is non-spherical, parts of the surface become subsonic at different times. Hence the choice of the reference point for measuring the cocoon expansion speed is important. Here we consider two locations, along the semi-minor (or transverse) and semi-major (or jet) axes of the cocoon. These are parts of the cocoon that become subsonic first (semi-minor) and last (semi-major axis). 

\begin{figure}
\begin{center}
\includegraphics[width=0.45\textwidth]{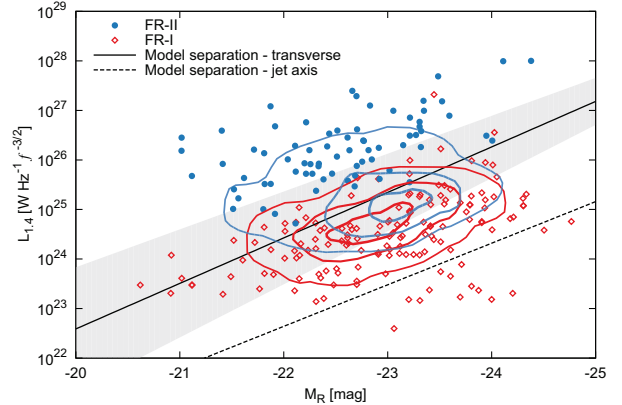} 
\end{center}
\caption[]{FR radio dichotomy as a function of $R$-band optical magnitude and $1.4 \rm\, GHz$  luminosity. Observations are from \citet[][points]{Owen+1994} and \citet[][contours]{Best+2009}. The contours correspond to locations with 16, 50 and 84\% fewer sources than the peak phase space density. {These observations are scaled assuming our modelled results used $f = 1$. Higher values in this uncertainty factor will cause these points to move to higher luminosities relative to our predictions.} The straight lines correspond to model predictions where half the sources are still expanding supersonically, and the other half subsonically. Calculations for expansion speed along the jet axis (dashed line) and transverse to the jet (solid) are shown, assuming a constant $10^7 \rm\, K$ temperature. The $1\sigma$ uncertainties due to variations in the temperature of the external medium are shown by the grey shading.}
\label{fig:FRmatrix3_A=4}
\end{figure}

We compare this simulated FR-I/II dichotomy with two sets of observations in Figure \ref{fig:FRmatrix3_A=4}. \citet{Owen+1994} reported the observed correlation between $R$-band optical magnitude, $1.4 \rm\, GHz$ radio luminosity and FR morphology for a sample of radio sources primarily found in Abell clusters. \citet{Best+2009} pointed out that, in this sample, the FR-Is are located predominantly at redshifts less than 0.1, with the bulk of the FR-II population being found at $z > 0.25$. By constructing a volume-limited sample, \citet{Best+2009} showed that the separation between FR-Is and FR-IIs in optical magnitude--radio luminosity phase space, albeit real, is much less clearcut than originally suggested by \citet{Owen+1994}.

In Figure \ref{fig:FRmatrix3_A=4}, the dashed and solid black lines represent fits from our model to the locations in jet power--stellar mass space where an equal fraction of the radio source active lifetime is spent in the subsonic (i.e. FR-I) and supersonic (i.e. FR-II) states. These fits are for a model using a constant $10^7 \rm\, K$ temperature for environments hosting our AGN, whilst the grey shading indicates the $1\sigma$ uncertainties corresponding to a factor of two variation in the temperature of the external medium. We convert stellar mass to $R$-band optical magnitude using the equations of \citet{Bell+2003}, and jet power to radio luminosity using our data. Based on our hypothesis, we expect sources lying above the semi-minor axis sonic transition line to be predominantly FR-IIs, with FR-Is lying below the jet axis line. The empirical separation between the observed FR-I and -II populations is consistent with the lines derived for the two subsonic points, and corresponds to a surface location somewhere between these extreme cases. Interestingly, the region where these two populations overlap in the \citet{Best+2009} sample is coincident with the parameter space in which our predicted model morphologies are strongly environment-dependent. {These conclusions are robust to realistic uncertainties in the jet power--luminosity relationship, e.g. $f \lesssim 5$.} We conclude that a comparison of the expansion speed of the cocoon to the local sound speed can thus successfully explain the observed FR-I/II dichotomy.

\begin{figure}
\begin{center}
\includegraphics[width=0.45\textwidth]{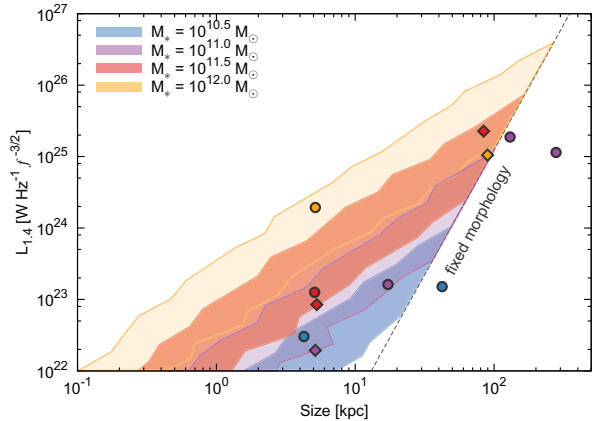}
\end{center}
\caption{FR radio morphology as a function of source size, $1.4 \rm\, GHz$ lobe luminosity and host galaxy stellar mass. {The shading maps regions in this phase space where our model predicts transonic phase evolution. Sources above these dividing lines for the relevant mass bin are predicted to be FR-IIs, while those below should be FR-Is. Objects lying within the shaded transonic regions at a given stellar mass can have either morphology.} The morphology of sources is not expected to change further upon entering the region beyond the grey dashed line. That is, a source retains its FR-II morphology if its $L\,$--$\,D$ track does not cross the morphological transition line before entering this region. {Note the overlap between shaded regions for different host galaxy masses.} Points show data from the CoNFIG sample of \citet{Gendre+2010}, with FR-IIs marked as circles and FR-Is as diamonds. Colours correspond to the stellar mass of the AGN host galaxy. {These observed points have been scaled using $f = 1$.}}
\label{fig:morphpred}
\end{figure}

Based on our findings, we tentatively identify radio sources as FR-Is or FR-IIs based on the Mach number modelled at the time of observation. Sources which are entirely supersonic are labelled FR-IIs, and those which are entirely subsonic are FR-Is. Sources which fall between these two categories could have either morphology, with the probability of being an FR-I increasing with the fraction of the cocoon surface which is subsonic. The FR-II population is predicted to consist of both smaller and younger sources than the FR-I population, irrespective of which temperature model is used, consistent with observations by \citet{Best+2007}. 

The locations in luminosity--size space in which FR-I and -IIs are thus expected to be found are plotted in Figure \ref{fig:morphpred} for four host galaxy stellar masses. {For each mass bin, the shading marks the regions in luminosity--size at which the expansion of radio sources is in the transonic phase (i.e. supersonic at semi-major but subsonic at semi-minor axis). The shaded regions include the uncertainty due to simulated variation in the host environment within each mass bin. At a given mass and size, radio sources in the supersonic expansion phase (which we identify with FR-IIs) have higher luminosities than these dividing lines, whilst sources expanding subsonically (our FR-Is) are less luminous. Sources lying within the shaded regions defined by the semi-major and semi-minor axis sonic transitions can have either FR morphology in our model.} Radio AGN inhabiting poor environments are more likely to be observed as FR-IIs than sources of the same size and luminosity in more massive hosts. However, the evolution of the larger sources is dictated mostly by the higher density environment at smaller radii. Sources which have reached sizes beyond the grey dotted line in Figure \ref{fig:morphpred} without entering the transonic phase are thus expected to retain their FR-II morphology. The morphology of the largest radio AGN is therefore somewhat sensitive to the initial conditions, in particular their host environment. We compare our radio morphology predictions with observations of the local extended radio source populations \citep{Gendre+2010}, and find the two to be in good agreement. That is, all their FR-IIs are in the supersonic or transonic expansion phase, whilst FR-Is are in the transonic or subsonic phase. {This result holds true for realistic values of the scaling factor $f$ in the jet power--luminosity relationship.}

\begin{figure}
\begin{center}
\includegraphics[width=0.45\textwidth]{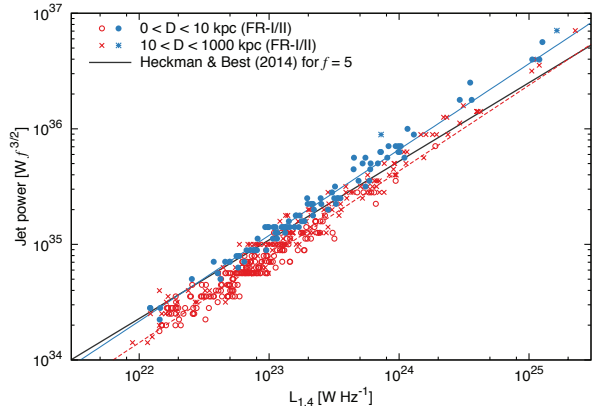}
\end{center}
\caption{Jet power as a function of the source size, luminosity and predicted Fanaroff-Riley morphology. {Radio AGN with sizes $D < 10 \rm\, kpc$ a marked by circles, whilst the largest sources ($D > 10 \rm\, kpc$) are shown with crosses. FR-Is are plotted with red, unfilled points and FR-IIs using blue, solid points.} Here, FR-IIs are identified with sources that are still expanding supersonically, while FR-Is are those sources for which at least part of the cocoon is expanding subsonically. Best fits to the smallest sources for both morphologies are plotted with dashed lines. {The cavities relation derived by \citet{Heckman+2014} is shown for comparison (grey), scaled using $f = 5$.}}
\label{fig:LQmorph}
\end{figure}

Finally, we note that our data can be used to derive an empirical relationship for the jet power based on the source morphology, size and luminosity observables. A number of loss processes associated with the synchrotron emitting electron populations render luminosity (by itself) a useful but flawed predictor of jet power, especially for evolved radio galaxies \citep{Shabala+2013}. We defer a detailed investigation of the jet power--radio luminosity relation to a future paper, but now examine its dependence on the Fanaroff-Riley morphology. The jet power (calculated in Section \ref{sec:Parameter estimation}) is plotted in Figure \ref{fig:LQmorph} as a function of size, luminosity and our tentative source morphology classification. 
{Also plotted is the cavity relationship of \citet{Heckman+2014}, scaled assuming our modelled results had a jet power--luminosity uncertainty factor of $f = 5$. That is, our model is in good agreement with other reported measurements of the jet power when assuming $f = 5$. This value agrees well with the findings of \citet{Daly+2012} using a different method of estimating jet power, and is well within the plausible range $f \sim 1$ - $20$ \citep[e.g.][]{Heckman+2014}.}

The jet power in Figure \ref{fig:LQmorph} shows a strong positive correlation with the $1.4 \rm\, GHz$ radio luminosity by itself, i.e. brighter sources tend to have higher jet powers. 
More of the variability in the jet power can be explained by considering the source size and morphology. The source size is plotted in Figure \ref{fig:LQmorph} for {two} size bins whilst the morphology is indicated by the point colour, with FR-Is in red and FR-IIs in blue. Larger sources tend to have higher jet powers than smaller sources of the same luminosity and morphology, consistent with the observations of \citet{Shabala+2013}. Further, radio AGN of FR-II morphology have jet powers approximately a factor of two higher than FR-Is of the same size and luminosity (dashed lines in Figure \ref{fig:LQmorph}). This is likely because at a given jet power, AGN expanding in poor environments are more likely to retain FR-II morphology; these sources will also be less luminous than their disrupted FR-I counterparts located in denser environments. Different empirical relationships will therefore apply for the two FR morphological classes, and thus radio AGN morphology must be considered in addition to size and luminosity in estimating jet power.

\section{AGN ENERGETICS AND FEEDBACK}
\label{sec:AGN ENERGETICS AND FEEDBACK}

\subsection{Jet powers and lifetimes}
\label{sec:Jet powers}

Having estimated AGN kinetic powers and lifetimes (Section \ref{sec:Parameter estimation}), we can now quantify the energetics of AGN feedback on galaxies. When drawing conclusions about the overall AGN population from our volume-limited sample we must be careful of selection effects though. Low jet power sources ($Q_{\rm tot} \lesssim 10^{34} {\rm\, W} f^{3/2}$) are theoretically undetectable by the FIRST/NVSS surveys at any time in their evolution, whilst higher jet power sources will not be detectable at later times when loss mechanisms (in particular the inverse Compton scattering of CMB photons) and Rayleigh-Taylor mixing cause the luminosity to fall rapidly. We quantify the fraction of undetected sources by modelling AGN with a range of jet powers, ages and environments. In our modelling, we assume that the AGN environments are well described by the scaled SAGE hot gas masses (Section \ref{sec:Radio AGN environments}), and we adopt the observed mass dependence of the radio-loud AGN fraction to quantify AGN intermittency. The figures in the following sections indicate this drop-out region with blue lines corresponding to detectable fractions of $84$, $50$ and $16\%$ of radio sources.

\subsubsection{Jet powers}

A combination of theoretical and empirical considerations suggest that the AGN jet power should to be largely independent of the host galaxy stellar mass. Jet generation models predict that the jet power increases approximately with the black hole accretion rate $\dot{M}_{\rm BH} \propto \dot{m} M_{\rm BH}$ \citep{Meier+2001}. The general consensus is that in massive ellipticals at low redshift, the active nucleus feeds energy back into its host galaxy at a rate balancing the loss of energy through radiative cooling \citep{Best+2007, McNamaraNulsen+2007, Fabian+2012}. Naively, we might expect this instantaneous accretion rate to balance the gas cooling rate, which scales with black hole mass (and thus stellar mass) as $\dot{M}_{\rm cool} \propto {M_{\rm BH}}^{1.5}$ \citep{Best+2005b}. The jet power would thus be expected to be proportional to the black hole and stellar masses as $Q_{\rm tot} \propto {M_{\rm BH}}^{1.5} \propto {M_\star}^{1.5}$. However, this jet power--stellar mass relationship needs to be modified for the AGN feedback cycle. Since there may be multiple AGN events within a single cooling time, we consider the time-averaged fuelling rate for the active nucleus, $f_{\rm RL} \dot{M}_{\rm BH}$, where $f_{\rm RL}$ is the radio-loud fraction. If the AGN are fuelled by the gravitational instability, $f_{\rm RL}$ is expected to scale with black hole mass as $f_{\rm RL} \propto {M_{\rm BH}}^{1.5}$ \citep{Pope+2012}. In heating-cooling equilibrium (i.e. $f_{\rm RL} \dot{M}_{\rm BH} \propto \dot{M}_{\rm cool}$) the jet power should therefore be related to the black hole and stellar masses as $Q_{\rm tot} \sim \dot{M}_{\rm cool}/f_{\rm RL} \propto {M_{\rm BH}}^{0} \propto {M_\star}^{0}$.

\begin{figure}
\begin{center}
\includegraphics[width=0.42\textwidth]{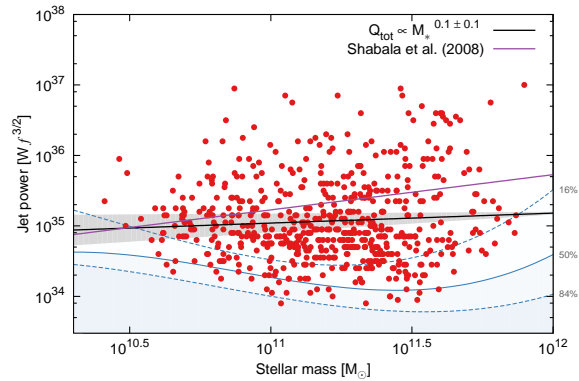} 
\end{center}
\caption{The correlation between jet power and observed stellar mass. Best fit is shown by a solid line, and uncertainties due to a factor of two variation in hot gas mass by the shaded region. The relationship obtained by \citet{Shabala+2008} using their simplified model for these same radio galaxies is shown in purple. The blue lines show the region of jet power--mass space where $16$, $50$ and $84\%$ of sources (from top to bottom) are expected to have luminosities  below the survey detection limits.}
\label{fig:QjetDistMass}
\end{figure}

The jet powers calculated for our sample are plotted as a function of stellar mass in Figure \ref{fig:QjetDistMass}. Unsurprisingly, a wide range of jet powers are observed at a given stellar mass. The relationship between these variables is analysed with a standard least squares linear regression (black line) yielding $Q_{\rm tot} \propto {M_\star}^{0.1\pm0.1}$. We also fit an L1 (least absolute errors) linear regression to provide a more robust estimate of the slope of $Q_{\rm tot} \propto {M_\star}^{0.0\pm0.1}$. By contrast, using their simpler FR-II only model \citet{Shabala+2008} found a steeper relationship of $Q_{\rm tot} \propto {M_\star}^{0.6\pm0.2}$, but with a comparable mean jet power (purple line).
Further, in this figure we show the region of jet power--mass space where sources are expected to have luminosities below the survey detection limits (blue lines and shading). Although detection limits may contribute to the lack of mass dependence for low jet powers ($Q_{\rm tot} < 10^{34.5} {\rm\, W} \, f^{3/2}$), the top end of the distribution corresponds to the real AGN population.
This relation shows no dependence on the host galaxy mass, in agreement with the theoretical expectation assuming long-term heating-cooling balance. Our results are therefore consistent with the maintenance nature of AGN feedback in the local Universe.

\subsubsection{Active and quiescent timescales}
\label{sec:Active and quiescent timescales}

In Section \ref{sec:ENVIRONMENT AND PARAMETER ESTIMATION} we derived the age of the radio source at the time of observation, not the total time over which the source is active. 
As sources age, they become less luminous due to a combination of synchrotron, inverse Compton and adiabatic losses. Because source size also increases with age, the surface brightness of the oldest sources drops rapidly, taking many of these close to (or below) the survey sensitivity threshold. Through the use of our marginalisation we may select very long-lived sources early in their active lifetime whilst they are detectable. If these sources were active up to their maximum detectable age we would expect the source ages to follow a uniform, linear distribution. However, more than 90\% of our sources are observed to be younger than half their maximum detectable age, with more than 70\% at a tenth of that age. There is therefore a real overabundance of young sources. We conclude that these young sources, and all other sources, must be observed somewhat close to their active age. The source active lifetime is thus taken to be double the observed age, i.e. $t_{\rm on} = 2 t_{\rm age}$, since the average source will be observed halfway through its evolution. 

\begin{figure}
\begin{center}
\includegraphics[width=0.42\textwidth]{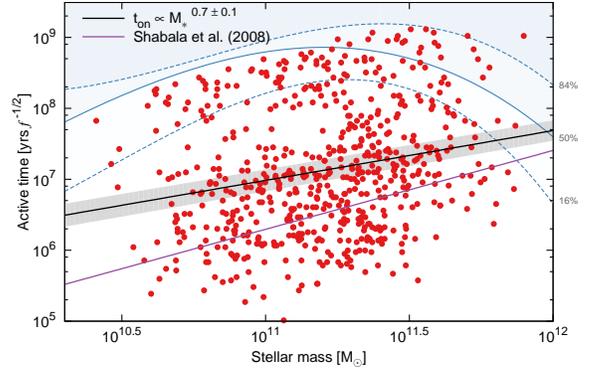} 
\end{center}
\caption{The active lifetime as a function of stellar mass. 
The blue lines are for the same fractions as in Figure \ref{fig:QjetDistMass} but with limits for both young and old sources.}
\label{fig:AgeDistMass}
\end{figure}

The relationship between the derived active lifetime and stellar mass for our radio AGN sample is shown in Figure \ref{fig:AgeDistMass}.
The mass dependence of the active lifetime is again calculated using a linear regression yielding $t_{\rm on} = (4.8\pm0.5 {\rm\,Myr} \, f^{-1/2}) (M_\star/10^{11} \rm\, M_\odot)^{0.7\pm0.1}$. The derived normalisation for our sample is higher than the $1 {\rm\,Myr} \, f^{-1/2}$ lifetime estimated by \citet{Shabala+2008} for a $10^{11} \rm\, M_\odot$ host galaxy, though the slope is comparable given their large uncertainties. Their younger ages result from the use of a simpler strong-shock limit supersonic expansion model for this mostly FR-I population.
Finally, the AGN active lifetimes should scale as $t_{\rm on} = f_{\rm RL} t_{\rm host}/N$, where $t_{\rm host}$ is the total age of the host galaxy group or cluster and $N$ is the number of AGN outbursts during this time \citep{Pope+2012}. The triggering frequency is thus related to the stellar mass as $\omega_{\rm trig} = N/t_{\rm host} = f_{\rm RL}/t_{\rm on} \propto {M_\star}^{0.8\pm0.1}$.
In general, the radio AGN in the most massive galaxies are therefore triggered more frequently, and are also active for longer in a given epoch.

\begin{figure}
\begin{center}
\includegraphics[width=0.42\textwidth]{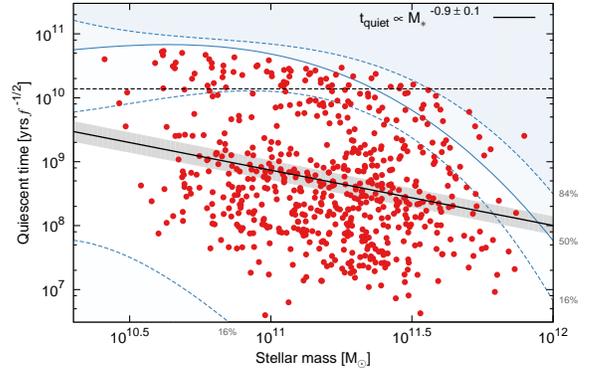}
\end{center}
\caption[]{The distribution of quiescent timescales plotted as a function of stellar mass, as for Figure \ref{fig:AgeDistMass}. The black dashed horizontal line is the Hubble time.}
\label{fig:QTSDistMass}
\end{figure}

The quiescent timescales (i.e. the time the AGN is inactive) are also relevant to understanding the AGN fuelling and triggering mechanisms.  The active lifetime can be used to determine the quiescent timescale as $t_{\rm quiet} = t_{\rm on}(1 - f_{\rm RL})/f_{\rm RL}$. Here we use the scaled relationship for the duty cycle of \citet{Best+2005b}. The quiescent timescales (plotted in Figure \ref{fig:QTSDistMass}) exhibit a strong negative correlation with stellar mass, driven largely by the strong dependence of $f_{\rm RL}$ on $M_\star$. Many of the radio sources hosted by galaxies with stellar masses less than $M_\star \approx 10^{11} \rm\, M_\odot$ are expected to be inactive for longer than the Hubble time. A large fraction of these low mass galaxies may therefore have never been in the active state. We note that here we used a source specific calculation of the active age, but a population average for the quiescent phase, which may not be applicable for individual sources.
We find the quiescent time scales with mass as $t_{\rm quiet} \propto {M_\star}^{-0.9\pm0.1}$. In the absence of a heating source, we would expect $t_{\rm quiet} \propto \dot{M}{_{\rm cool}}^{-1} \propto {M_\star}^{-1.5}$ \citep{Best+2005b}. 
The fact that the observed relationship is not as steep may suggest that feedback from powerful radio sources in massive galaxies plays an important role, flattening out the relationship at high masses as first proposed by \citet{Shabala+2008}. 

\subsubsection{AGN energetics and feedback}
\label{sec:AGN energetics and fuelling}

\begin{figure}
\begin{center}
\includegraphics[width=0.42\textwidth]{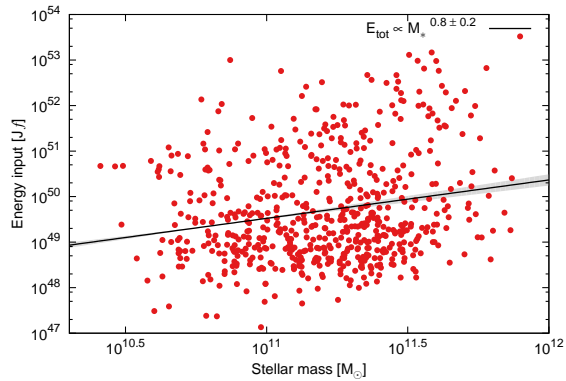}
\end{center}
\caption[]{The energy input of the AGN into their surroundings as a function of stellar mass. Detection limits are not shown since the effect of selection bias is similar for all masses, with approximately $30\%$ of sources undetected everywhere. While at a given stellar mass we observe a broad distribution in the total AGN energy output, there is a robust mass dependence for the average AGN.}
\label{fig:ESDistMass}
\end{figure}

Examining the energetics of AGN reveals which sources provide the most feedback in the low-redshift universe. The energy input by an AGN into its surroundings over an active phase is equal to the product of its jet power and the duration of the AGN event, $E_{\rm tot} = Q_{\rm tot} t_{\rm on}$. The input energies thus calculated for our sample are plotted in Figure \ref{fig:ESDistMass} as a function of stellar mass. The relationship between these variables is again analysed using a linear regression yielding $E_{\rm tot} \propto {M_\star}^{0.8 \pm 0.2}$. 
The energy injected by an AGN into its host in an outburst thus increases approximately linearly with the galaxy mass. Since the AGN in massive hosts are also triggered more frequently, we find the average energy injected by the central nucleus over time to scale even more steeply with mass as $\bar{E}_{\rm tot} \propto \omega_{\rm trig} E_{\rm tot} \propto {M_\star}^{1.6 \pm 0.2}$. This scaling is similar to the theoretical expectation for a cluster in heating-cooling equilibrium of $\bar{E}_{\rm tot} \propto \dot{M}_{\rm cool} \propto {M_\star}^{1.5}$.
The implications of these findings are discussed in more detail in Section \ref{sec:AGN heating and cluster cooling}. Massive hosts thus provide more feedback to their surroundings than their lower mass counterparts, as required by galaxy formation models \citep{Croton+2006,Bower+2006}. 

\begin{figure}
\begin{center}
\includegraphics[width=0.42\textwidth]{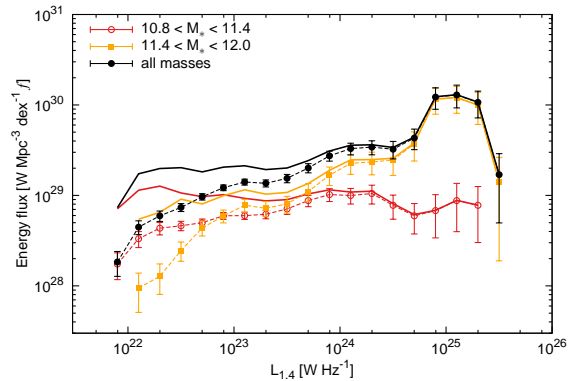}
\end{center}
\caption[]{The radio AGN energy flux (energy input by AGN into their hosts per unit volume) as a function of the observed $1.4 \rm\, GHz$ luminosity for two mass bins and our entire sample. We plot both the raw data (points and dashed lines) and also include curves that correct our data for survey selection effects (solid lines).}
\label{fig:EnergyFlux}
\end{figure}

We investigate which AGN provide the most feedback in the low-redshift universe by considering the number density of sources at a given stellar and radio luminosity, and their time averaged energy input, $\bar{E}_{\rm tot}$. The sources are binned in stellar mass and radio luminosity to yield luminosity functions (including $V/V_{\rm max}$ corrections) at a fixed stellar mass. The number densities in each bin are then weighted by the energy input of each radio AGN on a source-by-source basis. This energy flux (i.e. energy input per unit volume) is plotted as a function of radio luminosity in Figure \ref{fig:EnergyFlux} for two stellar mass bins and our entire sample. We plot the energy flux including corrections for the survey drop-out at low jet powers (solid lines) in addition to the uncorrected data (points and dashed lines). Poisson uncertainties ($1\sigma$ level) are plotted for the raw data.

The energy flux of radio AGN in high mass galaxies ($M_\star > 10^{11.4} \rm\, M_\odot$) increases strongly with luminosity until $L_{1.4} = 10^{25} \rm\, W\, Hz^{-1}$ where it peaks. These high luminosity sources contribute a factor of ten times more energy to the total galaxy population than those at $10^{23} \rm\,W \,Hz^{-1}$.  
By contrast, the energy flux of radio sources in lower mass hosts has only a very weak dependence on luminosity. These sources contribute a factor of four less energy to the low-redshift galaxy population than those in the most massive hosts, despite their greater number density. These result hold when corrections for the survey selection effects are applied. The lower energy flux associated with low luminosity radio sources and those in less massive hosts implies the amount of feedback energy such sources provide on the total galaxy population is much less than for the most massive galaxies near the $\sim 10^{25} \rm\, W\, Hz^{-1}$ luminosity peak.  Observing radio sources at even lower luminosities may therefore be unnecessary in further understanding the effect of AGN feedback on their host galaxies. Of course this does not take into account the very likely possibility of different coupling efficiencies with the gas for supersonic and subsonic cocoons. As discussed in Section \ref{sec:RADIO SOURCE MORPHOLOGY}, lower mass galaxies preferentially host supersonic phase FR-IIs whilst their more massive counterparts harbour subsonic FR-Is. Even for a given radio source morphology, numerical simulations of radio jets suggest the efficiency of feedback is strongly dependent on how the energy is supplied to the gas, rather than just the total amount \citep{Binney+2007}.

\subsection{AGN heating and cluster cooling}
\label{sec:AGN heating and cluster cooling}

In this section, we compare the time-averaged rate of heating from the AGN with the cooling in its host's hot environment. {Our jet powers calculated on a source-by-source basis provide a much broader distribution at a fixed stellar mass than previous work using empirical jet power--luminosity relationships \citep[e.g.][]{Best+2007}.}
We estimate the cooling rate, following the method of \citet{Best+2007}, by considering the fraction of X-ray luminosity arising from within the cooling radius, $r_{\rm cool}$. This rate is found as a function of the cluster velocity dispersion, $\sigma_{\rm v}$, by combining the results of \citet{Peres+1998} and \citet{Ortiz-Gil+2004}, $L_{\rm X} (r < r_{\rm cool}) \approx 1.34 \times 10^{37} (\sigma_{\rm v}/1000 \rm\, km\, s^{-1})^{4.1} \rm\, W$. Here, the cluster velocity dispersion is derived from the halo mass of each cluster using the relationship of \citet{Munari+2012}. 
The average rate of heating from the AGN is calculated using our results as the product of the jet power and the observed radio-loud fraction, i.e. $\bar{H} = Q_{\rm tot} f_{\rm RL} (M_\star)$. Again we assume the black hole-duty cycle relationship of \citet{Best+2005b}, but take $f_{\rm RL} = 1$ for the most massive black holes \citep[][and references therein]{Fabian+2012}. {Based on our findings in Figure \ref{fig:LQmorph}, we assume the uncertainty scaling factor in the jet power--luminosity relationship is $f \sim 5$ when comparing the heating and cooling rates.}

\begin{figure}
\begin{center}
\includegraphics[width=0.44\textwidth]{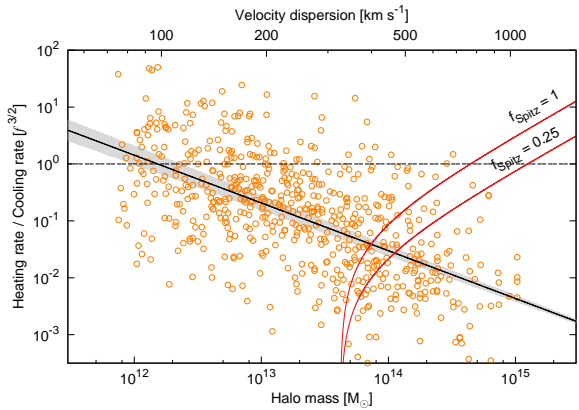}
\end{center}
\caption[]{The ratio of the AGN heating and gas cooling rates in the host cluster as a function of halo mass. The orange circles are the ratio of the time-averaged energy input from the jet and the cluster cooling rate. The solid black line is a best fit to this distribution with the grey shading showing uncertainties arising from our assumptions about the hot gas density. Heating due to thermal conduction is plotted in red for two scaling factors.}
\label{fig:Hrplot}
\end{figure}

The ratio of the time-averaged heating to the cluster cooling rate is plotted in Figure \ref{fig:Hrplot}. {Low mass galaxy haloes are consistent with a ratio of approximately unity (i.e. heating-cooling balance) for $f \lesssim 5$, however, high mass haloes ($>10^{14.5} \rm\, M_\odot$) require an additional heating mechanism. Such clusters have time-averaged heating rates a factor of $200 f^{-3/2}$ lower than the cluster cooling rates, though due to the large scatter in jet powers this difference is only significant at the $1\sigma$ level for $f \sim 5$. This discrepancy was previously reported by \citet{Best+2007}.}

Thermal conduction of energy from the reservoir of hot gas outside of the cooling radius into the cooler cluster centre may provide an additional source of heating in the most massive systems \citep[see][and references therein]{Best+2007}. The heating rate due to this thermal conduction was originally calculated for a pure hydrogen gas by \citet{Spitzer+1962}. Although this rate will be reduced in the presence of magnetic fields, \citet{Narayan+2001} argue that the suppression of thermal conduction may only be a factor of a few in the presence of turbulent magnetic fields. 
In Figure \ref{fig:Hrplot} we plot the effects of Spitzer conductivity for 25 and 100 per cent of the Spitzer value. The inclusion of this heating mechanism for realistic values of Spitzer conductivity can yield heating-cooling balance (within $1\sigma$ uncertainties) for all halo masses. That is, AGN feedback, in conjunction with thermal conduction, can maintain long term heating-cooling balance in even the most massive galaxy clusters, consistent with the findings of \citet{Best+2007} using a different jet power estimation method.

\section{FUELLING THE MONSTERS}
\label{sec:FUELLING THE MONSTERS}

In this section, we use our results to study black hole accretion flow mechanisms. This requires careful consideration of observational selection effects. 

\subsection{High- and low-excitation radio galaxies}
\label{sec:High- and low-excitation radio galaxies}

\begin{figure}
\begin{center}
\includegraphics[width=0.45\textwidth]{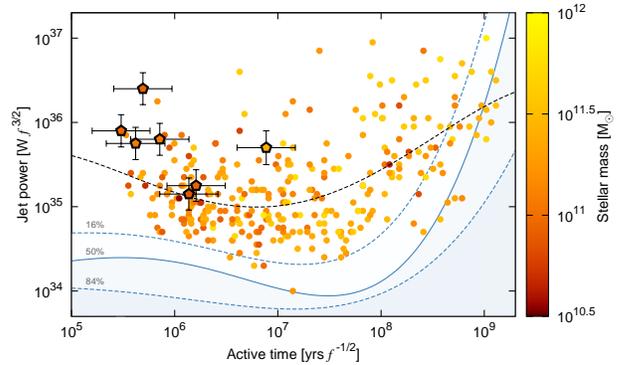}
\end{center}
\caption[]{The distribution of high- and low-excitation radio galaxies, pentagons and circles respectively, in active age-jet power space. The black dashed curve is a best fit to the distribution of LERGs, used to remove the effect of the active lifetime and selection effects from comparisons with the HERGs. The colouring of the points denotes the host galaxy stellar mass, whilst the blue lines indicate the times at which lower jet power sources are theoretically too faint to be detected (solid--50\% detectable, dashed--16 and 84\% detectable).}
\label{fig:Qtplot}
\end{figure}

First, we investigate whether the different black hole fuelling mechanisms of high- and low-excitation radio galaxies (HERGs and LERGs) correspond to differences in the active lifetime and jet power distributions, as might be expected if these objects probe different black hole accretion states \citep[e.g.][]{Hardcastle+2007, Best+2012}. There are not enough HERGs in our sample to draw reliable conclusions, however, the relative locations of the HERG and LERG sub-populations in active age--jet power space are shown in Figure \ref{fig:Qtplot}. The drop-out region of age--jet power space is marked with blue lines corresponding to detectable fractions of 84, 50 and 16\% as discussed in Section \ref{sec:Jet powers}. The observed ``correlation'' is thus largely a result of these survey selection effects, however the relative locations of HERGs and LERGs within this distribution are likely due to intrinsic differences. 
 
\begin{figure}
\begin{center}
\includegraphics[width=0.42\textwidth]{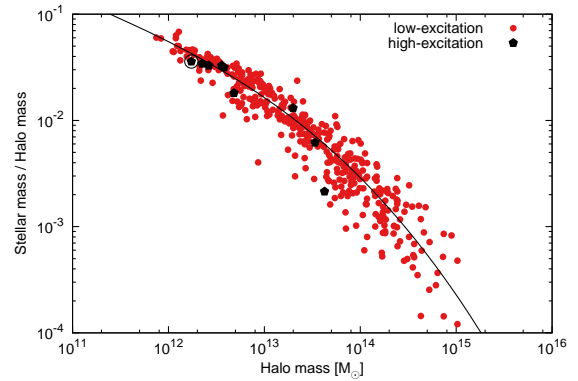}
\end{center}
\caption[]{Stellar-to-halo mass ratio of HERG (black pentagons) and LERG (red circles) populations as a function of halo mass. There is no statistically significant difference between the stellar-halo mass ratio for the HERGs and the best-fit line (black) for the LERG population. The single HERG identified as a spiral \citep[i.e. SDSS $r$-band $fracDeV$ parameter $<0.8$,][]{Padilla+2008} is circled, and again is consistent with the LERG distribution.}
\label{fig:halostellar}
\end{figure}

The HERG population has a mean stellar mass $0.24 \rm\, dex$ lower than for the LERGs, with this difference significant at the $3\sigma$ level. This is consistent with HERGs being fuelled by minor mergers, and LERGs through gas cooling which preferentially takes place in the most massive galaxies \citep{Hardcastle+2007, Best+2012, Shabala+2012}. 
The HERG and LERG radio sources, however, show no preference for inhabiting different environments for a given halo mass (Figure \ref{fig:halostellar}). Hence, we are not just seeing young HERG AGN because they are in poor environments (at a given $M_\star$).

The mean jet power of the HERG population is higher than that of the LERGs by $0.58 \rm\, dex$ ($3 \sigma$ significance), as shown in Figure \ref{fig:Qtplot}. This result holds even when the nine HERGs are compared with a sub-population of LERGs with the same distribution of stellar masses. These higher HERG jet powers are expected due to the higher black hole accretion rates in minor mergers.
Moreover, the mean active lifetime of the HERGs is $1.1 \rm\, dex$ lower than for the low-excitation population, with this difference significant at the $5\sigma$ level.
These younger HERG active ages likely result from these galaxies being associated with short lived epochs of high accretion arising from the merger of the host with a gas rich system \citep{Hardcastle+2007,Shabala+2012}. 
The timescales for the supply of cold gas in this scenario differ from the usual heating and cooling cycle of the IGM, which can fuel the lower accretion rate LERGs.

\subsection{Accretion rates and jet generation efficiencies}

The different fuelling modes of HERGs and LERGs are expected to yield different accretion rates.
We investigate the distribution of accretion rates for these populations by considering the kinetic power of the active nucleus relative to its Eddington luminosity, $L_{\rm Edd}$. The central black hole radiates energy directly from its accretion disk with the bolometric luminosity of the disk $L_{\rm rad}$ in addition to mechanical output of energy through the jets $L_{\rm mech} \equiv Q_{\rm tot}$. The power emitted by the AGN is thus related to the black hole mass accretion rate $\dot{M}_{\rm BH}$ as $(L_{\rm rad} + L_{\rm mech}) = \epsilon \dot{M}_{\rm BH} c^2$, where $\epsilon$ is the efficiency at which the accreted rest-mass energy is converted. The Eddington-scaled accretion rate is therefore

\begin{equation}
\dot{m} = \frac{\dot{M}_{\rm BH}}{\dot{M}_{\rm Edd}} = \frac{L_{\rm rad} + L_{\rm mech}}{L_{\rm Edd}}.
\label{accretion rate}
\end{equation}

{We note that this accretion rate is unaffected by realistic values of the jet power--luminosity scaling factor $f \lesssim 5$ since the mechanical jet power is much less than the disk bolometric luminosity.}
The bolometric radiative luminosity of each radio source is estimated using the observed luminosity of the [OIII] 5007 emission, and $L_{\rm rad} \approx 3500 L_{\rm OIII}$ \citep{Heckman+2004}. The uncertainty on individual estimates of the bolometric radiative luminosity is $\sim 0.4 \rm\, dex$. {Although 43\% of all radio sources in our sample have emission below the formal line flux uncertainties at the 3$\sigma$ level, this fraction falls to less than 10\% for objects classified as either HERGs or LERGs by \citet{Best+2012}. Following on from their work we adopt the $3\sigma$ upper limit to [O\,III] luminosity for such sources.}

In order to compare our accretion rate estimates with the theoretical expectations we need to correct for the survey selection effects. There are two factors we must consider: first, the drop-out of lower luminosity sources due to Rayleigh-Taylor mixing and late time loss mechanisms; and second, the greater likelihood of detecting long-lived sources from the total population compared to their younger counterparts. 
Specifically, the number density of sources in the population is probabilistically expected to be inversely proportional to its active age, $n_{\rm AGN}(t_{\rm on}) \propto 1/t_{\rm on}$.
The number of radio AGN missing due to the drop-out of low-luminosity sources is insignificant in comparison, as we discuss here. The luminosity of our simulated sources typically peaks at approximately $10 \rm\, Myrs$ (see Figure \ref{fig:pdtrack}). That is, we are able to observe all sources with jet powers above $10^{34.5} {\rm\, W} f^{3/2}$ in this age range, as seen in Figure \ref{fig:Qtplot}.
By contrast, all high jet power sources ($> 10^{35.5} {\rm\, W} f^{3/2}$) younger than $\sim 1 \rm\, Gyr$ can be detected with our survey. The distribution of counts of these high jet power sources in active lifetime space can thus be scaled to the source count at lower jet powers to predict the number of missing objects. In this manner, we find that at least half of the $10^{34.5} \leqslant Q_{\rm tot} < 10^{35.0} {\rm\, W} f^{3/2}$ jet power sources expected from the counts of $10^{35.5} \leqslant Q_{\rm tot} < 10^{36.0} {\rm\, W} f^{3/2}$ sources are observed. 
By contrast, correcting for the likelihood of observing objects based on their lifetimes yields a factor of ten increase in number density of radio sources for each order of magnitude decrease in the active age. Moreover, the derived active ages cover a much larger range (4 orders of magnitude) than the jet power ($2.5 \rm\, dex$). Hence, correcting observations for age bias is more important than the effects of different jet powers.

\begin{figure}
\begin{center}
\includegraphics[width=0.45\textwidth]{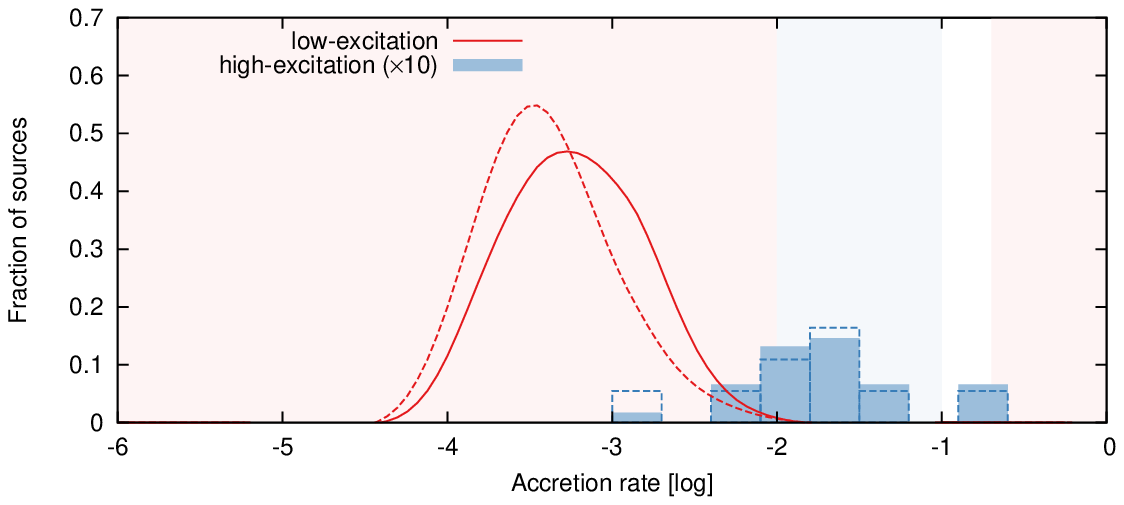} 
\text{}
\includegraphics[width=0.45\textwidth]{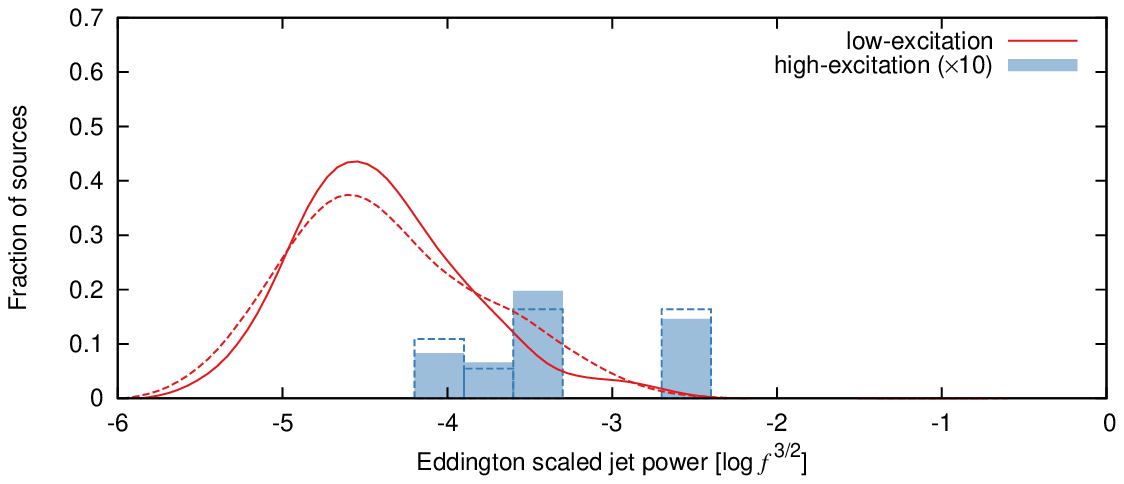} 
\text{}
\includegraphics[width=0.45\textwidth]{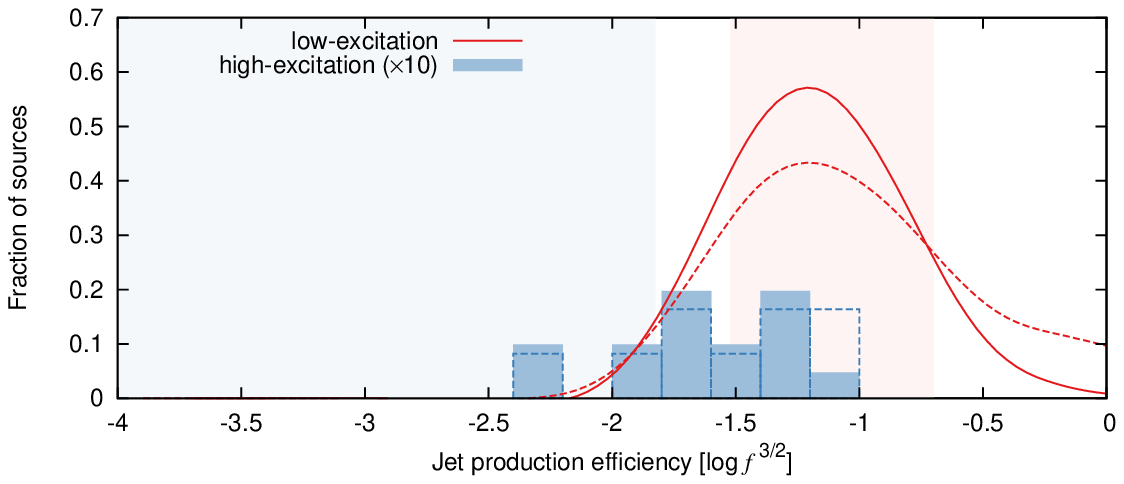} 
\end{center}
\caption[]{Top panel: distribution of Eddington-scaled accretion rates for LERGs (red) and HERGs (blue). The background shading shows the theoretically expected values for an ADAF (or slim disk; red) and a thin disk accretion flow (blue). The dashed lines show the uncorrected observed data, and solid lines are corrected for selection effects. Middle panel: mechanical jet power as a fraction of the Eddington luminosity. Bottom panel: efficiency of jet production. 
Fractions of the total AGN population are plotted, with HERGs scaled by a factor of ten to aid visibility.}
\label{fig:AccDist}
\end{figure}

The distribution of accretion rates thus corrected is now plotted in the top panel of Figure \ref{fig:AccDist} for the small population of HERGs (blue) and the much larger population of LERGs (red). The sub-population of LERGs has a mean accretion rate of $\dot{m} = 0.0006$ with a FWHM of $0.8 \rm\, dex$ with typical (i.e. $\gtrsim 95\%$ of AGN) values in the range $10^{-4} < \dot{m} < 10^{-2.5}$. The high-excitation population clearly has higher accretion rates with a mean of $\dot{m} = 0.02$, with a difference significant at the $9\sigma$ level. In fact, all but one of these nine objects have accretion rates greater than $\dot{m} = 10^{-2.5}$ compared to less than $2
\%$ of LERGs. Note that sources with luminosities below the flux detection limits of the FIRST and NVSS surveys for their entire lifetimes remain absent in these comparisons.
These accretion rate distributions are consistent with the analysis of \citet{Best+2012} for redshifts $z < 0.1$ (shown in their Figure 6), and largely with the theoretical division between the ADAF and thin disk accretion flow mechanisms at approximately $\dot{m} = 0.01$ \citep[red and blue background shading in Figure \ref{fig:AccDist};][]{Meier+2001}.

The middle panel of Figure \ref{fig:AccDist} shows the mechanical jet power scaled by the Eddington rate, $L_{\rm mech}/L_{\rm Edd}$. The HERGs have Eddington luminosity scaled jet powers greater than those of the LERGs by approximately a factor of twelve ($6\sigma$ significance). The relatively large fractional difference in accretion rates between the HERG and LERG populations compared to their Eddington-scaled jet powers indicates that the HERGs are much less efficient at producing jets. The jet production efficiency is plotted in the bottom panel of Figure \ref{fig:AccDist}. The LERG population has jet production efficiencies typically in the range $0.01 < \epsilon_{\rm jet, LERG} / f^{3/2} < 0.3$ with a mean efficiency of $\epsilon_{\rm jet, LERG} / f^{3/2} = 0.06$. By contrast, $82\%$ of HERGs have efficiencies less than this value with a mean of $\epsilon_{\rm jet, HERG} / f^{3/2} = 0.02$. The high-excitation radio galaxies are therefore approximately a factor of three less efficient at producing jets than the LERGs, and this difference is significant at the $4\sigma$ level.

\subsubsection{ADAFs}

The accretion rates and jet production efficiencies of the HERGs and LERGs can be compared to the predictions of \citet{Meier+2001} for black holes powered by an advection dominated or thin disk accretion flow. The jet production efficiency of an ADAF powered rotating black hole is related to its jet power $Q_{\rm tot}$ \citep[see Equation 7 of][]{Meier+2001} and the Eddington-scaled accretion rate $\dot{m}$ by

\begin{equation}
\epsilon_{\rm ADAF} / f^{3/2} = \frac{Q_{\rm ADAF}/L_{\rm Edd}}{\dot{m}} = 0.05 (0.55 + 1.5 j + j^2) .
\end{equation}

The black hole spin $j \in [0,1]$ is a difficult quantity to measure. \citet{MR+2011} have estimated that in the low-redshift universe the mean black hole spin is $\left<j \right> \sim 0.35$, with approximately $24\%$ of black holes having spins greater than $j = 0.5$. Variation in this parameter can generate jet production efficiencies in the range $0.03 < \epsilon_{\rm ADAF} / f^{3/2} < 0.2$, consistent with the peak in the drop-out corrected LERG distribution (solid red line; Figure \ref{fig:AccDist}, bottom) which we identify with ADAFs. The smaller ``bump'' in the observed distribution (dashed red line) at an efficiency of approximately $\epsilon_{\rm jet, LERG} / f^{3/2} \geqslant 0.5$ is not consistent with this theoretical model. These sources have median radio luminosities a factor of 45 greater than the rest of the LERG population, whilst sizes, jet powers and active lifetimes are 25 times larger. The typical sizes of these sources' radio cocoons are of the order of $100 \rm\, kpc$ with a median active lifetime of $100 {\rm\,Myr} \, f^{-1/2}$. The transit time of plasma along their jets leads to a lag of order $10^6$ to $10^7 \rm\, years$ between changes in the central engine being reflected in the radio lobe emission. The observed radio luminosity from the cocoon may therefore reflect an earlier time with higher accretion rates than the present. We speculate that these objects predicted by our model may be ``dying'' quasars.

\subsubsection{Thin disks}

The jet production efficiency of a thin disk accretion flow AGN (observationally identified with a HERG) is similarly derived using Equation 5 of \citet{Meier+2001}. That is,

\begin{equation}
\epsilon_{\rm TD} / f^{3/2} = 0.005 (1 + 1.1 j + 0.29 j^2) \left(\frac{M_{\rm BH}}{10^9 \rm\, M_\odot} \right)^{-0.1} \dot{m}^{0.2} .
\end{equation}

The highest efficiency that can be produced by this mechanism for the accretion rates of our sample is $\epsilon_{\rm TD} / f^{3/2} = 0.015$. Two of the nine HERGs in our sample have lower jet production efficiencies, however, seven have efficiencies exceeding this value rising up to $\epsilon_{\rm jet, HERG} / f^{3/2} = 0.09$.
These observations may be explained by a ``combined disk'' accretion flow for $\dot{m} \lesssim 0.01$ \citep[e.g.][noting that the high efficiency sources have such accretion rates]{Narayan+1996, Pu+2012}. These accretion flows have a geometrically thin outer disk but a puffed up inner region like an ADAF. Black holes fuelled by such accretion flows could produce jets moderately efficiently due to the ADAF core but still exhibit the strong emission lines of HERGs.

\section{CONCLUSIONS}
\label{sec:CONCLUSIONS}

We presented a new dynamical model for jet-driven radio source evolution. This model differs in three respects from previous models: (1) it includes both the supersonic and subsonic cocoon expansion phases, (2) uses external pressure profiles based on semi-analytic galaxy formation models and X-ray observations, and (3) includes the Rayleigh-Taylor mixing of the cocoon with the surrounding material. 
We applied our model to a low-redshift volume-limited sample of radio AGN, and estimated the jet powers and active lifetimes of these objects. Radio sources in massive galaxies were found to remain active for longer, spend less time in the quiescent phase, and inject more energy into their hosts than their less massive counterparts. {The majority of AGN energy output in the low-redshift universe is provided by sources in massive hosts with $\sim 10^{25} \rm\, W\, Hz^{-1}$ luminosities.} The relationship between the AGN jet powers and the stellar masses of their host galaxies, $Q_{\rm tot} \propto {M_\star}^{0.1\pm0.1}$, is in agreement with the theoretically expected (lack of) dependence for maintenance-mode AGN feedback. The host haloes of these AGN are likely to be in or close to long-term heating-cooling balance, when considering both thermal conduction and the suppression of gas cooling by the AGN.
Our dynamical model can also explain the observed FR-I/II radio morphology dichotomy.

High-excitation radio galaxies were found to have active lifetimes on average a factor of twelve lower and jet powers a factor of four higher than the low-excitation radio galaxies. This is consistent with the sporadic nature of their merger-driven cold gas accretion. The bimodality in LERG/HERG accretion rates is consistent with the divide between the ADAF and thin disk accretion flow mechanisms at approximately $\dot{m} = 0.01$. The high-excitation radio galaxies in our sample were found to be a factor of three times less efficient at producing jets than the LERGs. The derived jet production efficiencies are broadly consistent with the ranges allowed by radiatively efficient and inefficient flow jet production models. The jet production efficiency of the lowest accretion rate HERGs exceeds the maximum predicted for a thin disk accretion flow, which we suggest provides support for the ``combined disk'' accretion flow model.

The improved characterisation of the interaction between AGN and their host galaxies resulting from this work should provide valuable input into galaxy formation and evolution models. This model will also be a useful tool in the interpretation of high redshift survey data, especially for \emph{Square Kilometre Array} (SKA) pathfinder surveys. An important development will be extending the model to estimate source ages from observed spectral indices, since the resolution of the radio cocoons may not be possible except for the largest sources (e.g. $80 \rm\, kpc$ at $z \sim 1$ for ASKAP EMU). At lower redshifts, future sensitive X-ray surveys of the gas density and temperature profiles of the AGN environments (e.g. ATHENA\footnote{\url{http://www.the-athena-x-ray-observatory.eu/}}) will be complementary to our work.

R.T. thanks the University of Tasmania for a University Club Honours Scholarship and an Elite Research Scholarship. S.S. thanks the Australian Research Council for an Early Career Fellowship, DE130101399. We are grateful to Darren Croton for useful discussions, and the referee for multiple insightful comments which have significantly improved the manuscript.

\begin{appendix}
\section{Ellipsoidal Trigonometry}

The working surface pressure at each angle $\theta$ differs from that along the jet axis due to the non-spherical nature of the cocoon.
There are two relevant pressures here: the external pressure due to the ambient medium, and ram pressure due to the cocoon expansion. The latter requires a conversion from the radial expansion rate of each cocoon element to the velocity normal to its surface. Sections of the cocoon expanding supersonically will maintain self-similar growth (approximately self-similar when outside strong-shock limit) and can thus be modelled by an ellipsoid with a constant axis ratio $A$. The component of the velocity normal to the surface at angle $\theta$ is related to the expansion rate along the jet axis $v_\perp(\theta = 0)$ through

\begin{equation}
\zeta(\theta) = \frac{v_\perp(\theta)}{v_\perp(\theta = 0)} = \left[\frac{A^2 \sin^2 \theta + \cos^2 \theta}{A^4 \sin^2 \theta + \cos^2 \theta} \right]^{1/2} ,
\end{equation}

where we have defined a dimensionless velocity $\zeta$. This coefficient can be calculated with a typical value of the cocoon axis ratio for powerful FR-II sources. We now need to relate this normal velocity along the jet axis, $v_\perp(\theta = 0)$, to the rate of cocoon expansion in the $\theta$ direction for the assumed ellipsoidal cocoon. The radial distance to the working surface at angle $\theta$ from the central black hole is related to the radius along the jet axis $R(\theta = 0)$ by

\begin{equation}
\eta(\theta) = \frac{R(\theta)}{R(\theta = 0)} = \frac{\dot{R}(\theta)}{v_\perp(\theta = 0)} = \frac{1}{\sqrt{A^2 \sin^2 \theta + \cos^2 \theta}} ,
\end{equation}

where we have defined a dimensionless radius $\eta$. The velocity normal to the surface at each angle $\theta$ is therefore related to the radial expansion rate of the relevant cocoon volume element through $v_\perp = \zeta \dot{R}/\eta$.

These definitions for the dimensionless radius and velocity are also useful in describing the initial self-similar growth phase of the entire cocoon. Here, these equations allow for analytically tractable solutions for the expansion of each volume element and thus the fraction of the jet power required by each element to maintain self-similarity. In the strong-shock supersonic limit we find the jet power is proportional to the angular location of the volume element as

\begin{equation}
\frac{8\pi k \sin\theta d\theta}{3 (\Gamma_{\rm x} + 1)} \left[(3 \Gamma_{\rm c} - \beta)R^{2 - \beta} \dot{R}^3 + 2R^{3 - \beta} \dot{R} \ddot{R} \right]_{\theta = 0} \eta^{3 - \beta}(\theta) \zeta^2(\theta) = (\Gamma_{\rm c} - 1) Q\, d\lambda(\theta) ,
\end{equation}

using Equations \ref{delta volume}, \ref{hotspot pressure 4} and \ref{adiabatic expansion}.

\end{appendix}

\end{document}